\documentclass[12pt]{article}
\usepackage{graphicx} 
\usepackage{amsmath}

\usepackage{adjustbox}
\usepackage{makecell}
\usepackage{rotating}
\usepackage{amssymb}
\usepackage{arydshln}
\usepackage{arxiv}

\usepackage[utf8]{inputenc} 
\usepackage[T1]{fontenc}    
\usepackage{hyperref}       
\usepackage{url}            
\usepackage{booktabs}       
\usepackage{amsfonts}       
\usepackage{nicefrac}       
\usepackage{microtype}      
\usepackage{cleveref}       
\usepackage{natbib}
\usepackage{doi}

\title{A generalized multiple-intervention stepped wedge design framework for treatment effect estimation in the presence of non-uniform cluster-period correlation structures}

\author{ \href{https://orcid.org/0000-0002-8663-5079}{\includegraphics[scale=0.06]{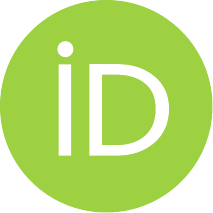}\hspace{1mm}Samantha M.~Levy, PhD}\\
	Asymmetric Operations Sector\\
	The Johns Hopkins University Applied Physics Laboratory\\
	Laurel, Maryland \\
	\texttt{samantha.levy@jhuapl.edu} \\
	\And
	\href{https://orcid.org/0000-0003-2505-0090}{\includegraphics[scale=0.06]{orcid.pdf}\hspace{1mm} Jose-Miguel~Yamal, PhD} \\
	Department of Biostatistics and Data Science\\
 UTHealth Houston School of Public Health, The University of Texas Health Science Center at Houston\\
	Houston, Texas \\
	\texttt{Jose-Miguel.Yamal@uth.tmc.edu} }

\renewcommand{\shorttitle}{Non-uniform correlation and power in multiple-intervention SWDs}

\hypersetup{
pdftitle={Non-uniform correlation and power in multiple-intervention SWDs},
pdfsubject={stat.ME},
pdfauthor={Samantha M.~Levy, Jose-Miguel~Yamal},
pdfkeywords={stepped wedge design, multiple intervention, cluster randomized trials, intracluster correlation, correlation misspecification, power analysis},
}

\begin{document}
\maketitle

\begin{abstract}
	Existing power and design methods for multiple-intervention stepped wedge designs (M-SWDs) typically assume exchangeable cluster-period correlation, despite evidence that correlation often decays over time. Misspecification of this correlation structure can substantially distort variance estimation and power, particularly for treatment interaction effects. We develop a unified covariance framework for M-SWDs that separates intracluster correlation from an explicit cluster-period correlation matrix. This formulation accommodates exchangeable, autoregressive, and more general distance-dependent correlation structures while preserving closed-form expressions for the variance of treatment effect estimators under linear mixed models. Using analytic results and simulation studies, we demonstrate that assuming uniform correlation when the true structure is time-dependent can lead to substantial power mischaracterization. Specifically, we find that designs calibrated under independence assumptions may be overly conservative and compound symmetry can be either optimistic or conservative. These findings demonstrate the importance of explicitly modeling cluster-period correlation at the design stage of M-SWDs and provide practical guidance for power calculation and design selection in realistic settings. 
\end{abstract}


\section{Background}
Large-scale implementation and health system-wide trials often require designs that balance methodological rigor with logistical and ethical feasibility. When interventions are intended to eventually be delivered to all clusters - such as system-wide policies, educational programs, or public health rollouts - the stepped wedge design (SWD) provides a pragmatic and statistically efficient framework for phased implementation. SWDs are a type of cluster randomized trial in which each cluster, or group of clusters, transitions from a control arm to an intervention arm in a randomized, staggered sequence \cite{the_gambia_hepatitis_study_group_gambia_1987, hussey_design_2007}. Due to their practical nature and logistical flexibility, SWDs are increasingly used across a variety of applications \cite{brown_stepped_2006, mdege_systematic_2011}. Specifically, the flexibility to make both within- and between-cluster comparisons over time allows for significant advantages in pragmatic settings especially when universal roll-out of the intervention is planned or ethically mandated. 

Early analytic frameworks for SWDs, such as the Hussey-Hughes model, assume a single intervention, a single control, one intra-cluster correlation (ICC), and few random-effect components that inadequately account for the complexity of real-world trials \cite{hussey_design_2007, grayling_stepped_2017}. Recent work has shown that assuming uniform correlation in stepped wedge designs, despite the presence of repeated observations within clusters over time, can substantially mischaracterize variance and power, motivating more flexible correlation structures  \cite{thompson_bias_2017, nickless_2018,  vodal_2022_missp, hemming_modelinghetero_2018, murray_intraclass_1995, donner_design_1982, hooper_sample_2016, kasza_impact_2019, girling_statistical_2016, li_mixed-effects_2021}.

\begin{figure}
    \centering
    \includegraphics[width=0.6\linewidth]{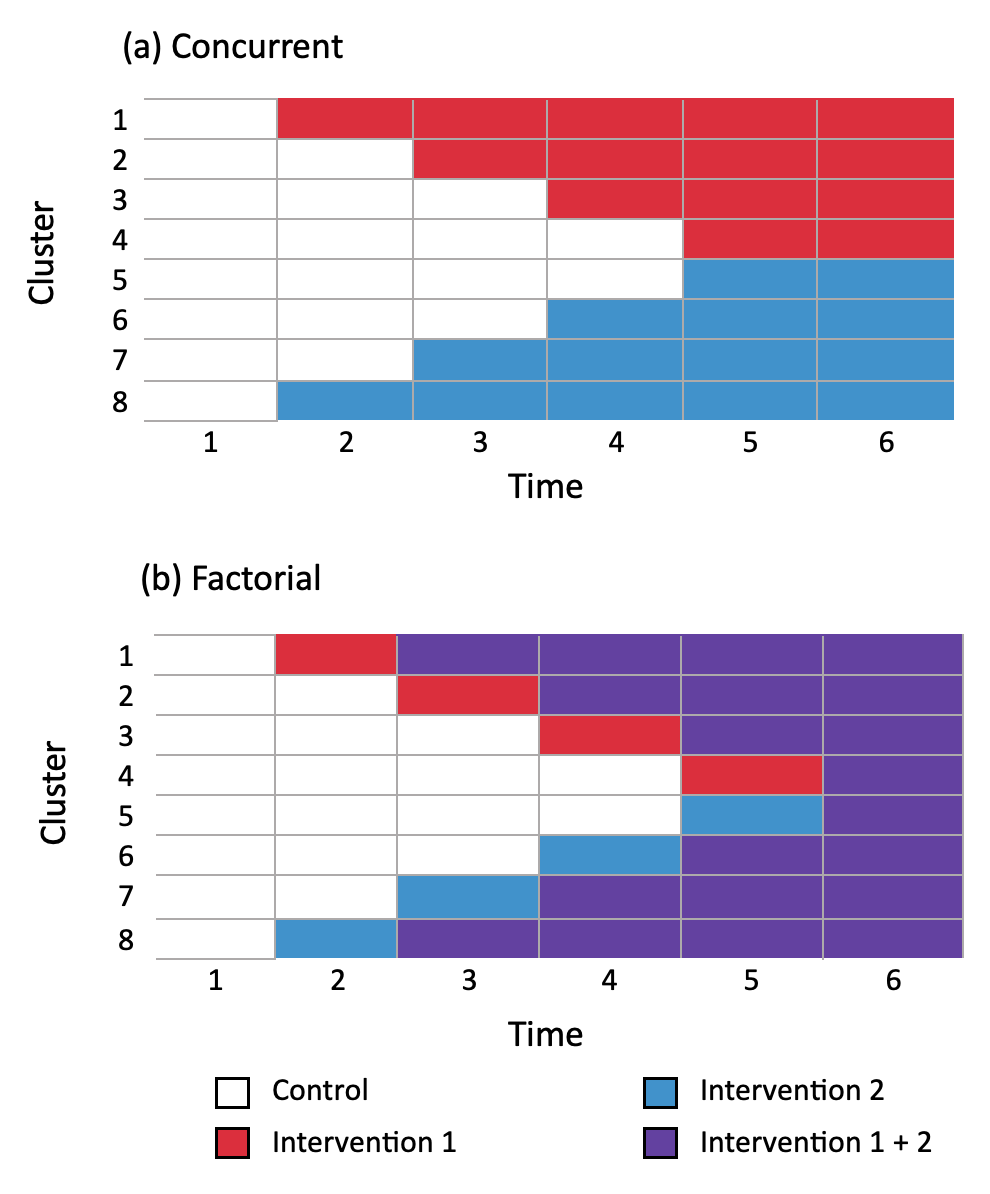}
    \caption[Aim 1 Concurrent and Factorial Design]{Sample (a) Concurrent and (b) Factorial multiple-intervention stepped wedge design studies with 8 clusters and 6 periods}
    \label{fig:concurfact}
\end{figure}

A growing area in SWD research and application is the expansion to multiple-intervention stepped wedge design (M-SWD) studies. Formally proposed by Lyons et al., M-SWDs involve at least two interventions, plus a control, with or without an interaction term, in a flexible, staggered rollout \cite{lyons_proposed_2017}. Two common types of M-SWDs are concurrent and factorial, as shown in Figure \ref{fig:concurfact}. Due to the pragmatic advantages, M-SWDs studies have been increasingly implemented across public health applications; however statistical methods to support their design and analysis remain limited \cite{pol_effectiveness_2017, pol_effectiveness_2019, finney_rutten_evaluating_2018, zhu_enhancing_2023, Zhang2025-ir}.  

The linear mixed model (LMM) framework proposed by Sundin and Crespi \cite{sundin_power_2022} represents an important step toward power analysis for M-SWDs.  However, existing approaches have assumed uniform or exchangeable correlation structures that may not capture realistic temporal dependencies within clusters. In M-SWDs, this assumption is particularly restrictive as treatment contrasts often span non-consecutive periods and correlation may decay over time. 

To address these limitations, we developed a generalized multiple-intervention stepped wedge design framework that allows structured within-cluster correlation matrices, as well as unified ICCs and cluster-period correlations to accommodate non-uniform, time-dependent correlation. We also provide analytic expressions towards variance, power, and efficiency under these structures. In Section \S\ref{sec:methaim1} we extend the existing framework to a general structure under a flexible correlation matrix, examine common correlation structures in M-SWD, and frame the variance and parameter estimation in the context of non-uniform correlation structures. In Section \S\ref{sec:sim} we design an analytic and simulation study to validate and assess these results. In Section \S\ref{sec:aim1results} we present the results of this study, showing the impact of these correlation structures and the impact of mis-specifying the correlation structure on precision and bias. We conclude with a discussion in Section \S\ref{sec:discussion1} of the impact of our work including limitations and potential future work.

\section{Methodology}\label{sec:methaim1}

\subsection{Notation and Base Model Description}\label{sect:Aim1notation}

In this section, we start with the established model for a stepped wedge design with a single intervention and uniform correlation structure. We then build up to more complex correlation structures and add flexibility for multiple interventions and an interaction term. The "Hussey-Hughes" parameterization of the stepped wedge design\cite{hussey_design_2007} for an individual level response $Y_{ijk}$, where $i = 1, \ldots, I$ clusters, $j=1, \ldots, T$ time points, $k=1, \ldots, N$ individuals per cluster can be written as

\begin{equation}\label{eq:HH2}
Y_{ijk} = \mu  + \beta_j + \alpha_i  + \psi_{ik} + \theta X_{ij} + e_{ijk}
\end{equation}
where $X_{ij}$ is a \{0,1\} indicator of whether cluster $i$ received the intervention at time point $j$, $\mu$ is a fixed intercept, $\beta_j$ is a fixed effect for time period $j$, such that for identifiability $\beta_T = 0$, $\alpha_i$ is the random effect for cluster $i$ such that $\alpha_i \sim N(0, \sigma^2_\alpha)$, $\psi_{ik}$ is the random effect for individual $k$ in cluster $i$ such that $\psi_{ik} \sim N(0,\sigma^2_\psi)$, and $e_{ijk} \stackrel{iid}{\sim} N(0, \sigma^2_e)$ is the marginal variance. We define this model as an independent correlation structure. This baseline formula assumes independence across time within clusters, which is often unrealistic in stepped wedge designs where outcomes are repeatedly measured within clusters over time.

To allow for additional within-cluster dependence across time periods, we extend the base model by introducing a cluster-period random effect such that

\begin{equation}\label{eq:fullsing}
Y_{ijk}= \mu  + \beta_j  + \alpha_i + \nu_{ij} + \psi_{ik}  + \theta X_{ij} + e_{ijk} 
\end{equation} 
where all terms are as defined above and $\nu_{ij} \sim N(0, \sigma^2_\nu)$ is the random effect for cluster $i$ at time $j$.  The total variance of an individual response, $Y_{ijk}$, is thus 
\begin{equation}\label{eq:sigma2}
\text{Var}(Y_{ijk})= \sigma^2_y = \text{Var}(\alpha_i) + \text{Var}(\nu_{ij})+  \text{Var}(\psi_{ik}) + \text{Var}(e_{ijk}) = \sigma^2_\alpha + \sigma^2_\nu+ \sigma^2_\psi + \sigma^2_e.   
\end{equation}
We define this as an exchangeable correlation structure. For repeated cross-sectional designs, $\psi_{ik} = 0$. This variance decomposition motivates correlation formulations that distinguish dependence within a period from dependence across periods. 

When interested in looking at more than one treatment effect, such as the comparison of two different protocols to a standard of care, we expand this model to include $Q$ interventions \cite{lyons_proposed_2017, sundin_power_2022}. Assuming additive treatment effects, the SWD model for $Q$ interventions can be written as

\begin{equation} \label{eq:fullestsimp}
    Y_{ijk} = \mu + \beta_j + \alpha_i + \nu_{ij} + \psi_{ik} + \sum_{q=1}^Q \theta_q X_{qij} + e_{ijk}
\end{equation}
where $\theta_q$ is the fixed treatment effect for treatment $q$ and $X_{qij}$ is a \{0,1\} indicator of whether cluster $i$ receives treatment $q$ at time $j$. This model reduces to the standard one treatment model when $Q=1$.  

A common use case of multiple-intervention SWD involves two interventions delivered alone and in combination, motivating the inclusion of an interaction term. This model would be represented as
\begin{equation}\label{eq:fullmult}
    Y_{ijk} = \mu + \beta_j + \alpha_i + \nu_{ij} + \psi_{ik} + \theta_1 X_{1ij} + \theta_2 X_{2ij} + \theta_3 X_{1ij}X_{2ij} + e_{ijk}.
\end{equation}

\subsection{Intra-Cluster Correlations Under an Exchangeable Correlation Structure}\label{sec:genICC}

Under the exchangeable baseline model, we summarize the within-cluster dependence using the within-period ICC (WPICC), 
    \begin{equation}
    \rho_w =  \text{Corr}(Y_{ijk}, Y_{ijl})= \frac{\sigma^2_\alpha + \sigma^2_\nu}{\sigma^2_y}
    \end{equation}
for $k \neq l$ and the between-period ICC (BPICC), 
    \begin{equation}
    \rho_b =  \text{Corr}(Y_{ijk}, Y_{iml}) = \frac{\sigma^2_\alpha}{\sigma^2_y}
    \end{equation}
for $k \neq l$ and $j \neq m$. The ICCs summarize dependence at the individual level.

\subsection{Cluster-Period Means and Variances Under an Exchangeable Correlation Structure}
While the ICCs summarize individual level dependence, estimation in stepped wedge design clinical trials is typically driven by the covariance structure of the cluster-period means. For a given cluster $i$ at time $j$, we define the cluster-period mean as
\begin{equation*}
    \bar{Y}_{ij.} = \frac{1}{N} \sum_{k=1}^N Y_{ijk} .
\end{equation*}
Under this exchangeable correlation structure, the model for cluster-period mean is 

\begin{equation*}
    \bar{Y}_{ij.}= \mu + \beta_j +  \alpha_i + \nu_{ij} + \psi_{i.} + \sum_{q=1}^Q \theta_q X_{qij} + e_{ij.}
\end{equation*}
where $\psi_{i. }= \frac{1}{N}\sum_{k=1}^N\psi_{ik} \sim N(0, \sigma^2_\psi/N = \sigma_\zeta^2)$ is the average subject-level effect within a cluster-period and $e_{ij.} = \frac{1}{N}\sum_{k=1}^Ne_{ijk} \sim N(0, \sigma^2_e/N = \sigma^2_c)$ is the average residual within a cluster-period. 

We then define the total variance of the cluster-period mean as
\begin{equation} \label{eq:eta}
    \eta = \text{Var}(\bar{Y}_{ij.}) = \sigma^2_\alpha + \sigma^2_\nu + \sigma^2_\zeta + \sigma^2_c
\end{equation}
where $\sigma^2_\alpha$ is the persistent cluster-level variance, $\sigma^2_\zeta$ is the cluster-level aggregate of subject variance, $\sigma^2_\nu$ is the cluster-period variance, and $\sigma^2_c$ is the cluster-period mean level residual error. $\eta$ characterizes the marginal variability of a single cluster-period mean.

\subsection{Cluster-Period Correlation and the $\omega$ Framework}
Efficiency in stepped wedge designs depends not only on the marginal variability, $\eta$, but also on how information is shared across periods within a cluster. We define the cluster-period correlation, $\omega$, as the proportion of total cluster-period variance that is shared over time. Under an exchangeable correlation structure modeled in equations \eqref{eq:fullsing} and \eqref{eq:fullmult}, the correlation between two time points $j$ and $m$ for the same cluster is given by
\begin{equation} \label{eq:omegaex}
    \omega_{jm} = \text{Corr}(\bar{Y}_{ij.}, \bar{Y}_{im.}) = \frac{\sigma^2_\alpha + \sigma^2_\zeta}{\eta}
\end{equation}
for $j \neq m$ as $\sigma^2_\nu$ and $\sigma^2_c$ are assumed independent across time under the exchangeable correlation structure, $\sigma^2_\alpha$ and $\sigma^2_\zeta$ represent the variance components that are shared across time within a cluster. As a result, the correlation between any two cluster-period means is constant for all $j \neq m$ and we can simplify to $\omega_{jm} = \omega$ under the exchangeable and uniform correlation structures.

\subsection{Covariance Matrix Under Exchangeable Structure}\label{sect:covbasic}
We can expand this correlation into a full covariance matrix for the cluster-period means under an exchangeable correlation such that for cluster $i$,
\begin{equation}\label{eq:Vex}
    \text{Cov}(\bar{Y}_i)=V=(\sigma^2_\nu + \sigma^2_c)I_T + (\sigma^2_\alpha + \sigma^2_\zeta)J_T
\end{equation}
where $I_T$ is the $T \times T$ identity matrix and $J_T$ is the $T \times T$ matrix of ones. Under this parameterization, $\sigma^2_\alpha + \sigma^2_\zeta$ is grouped as the shared variance across time while $\sigma^2_\nu + \sigma^2_c$ is the time-specific variance. We can reparameterize the covariance matrix in terms of the total cluster-period variance, $\eta$, and the constant cluster-period correlation, $\omega$. Since $\eta =\sigma^2_\alpha + \sigma^2_\nu + \sigma^2_\zeta + \sigma^2_c$ and $\omega = \frac{\sigma^2_\alpha + \sigma^2_c}{\eta}= \frac{\sigma^2_\alpha + \sigma^2_c}{\sigma^2_\alpha + \sigma^2_\nu + \sigma^2_\zeta + \sigma^2_c}$, we find that $\eta\omega  = \sigma^2_\alpha + \sigma^2_\zeta$. Similarly, $\eta(1-\omega) = \sigma^2_\nu + \sigma^2_\zeta$. Thus we can say 
\begin{equation}\label{eq:Vexeta}
    \text{Cov}(\bar{Y}_i)=V = (\sigma^2_\nu + \sigma^2_c)I_T + (\sigma^2_\alpha + \sigma^2_\zeta)J_T = \eta\omega J_T + \eta(1-\omega)I_T=\eta(\omega J_T+(1-\omega)I_T).
\end{equation}

\subsection{Flexible Correlation Structure with the $R$ matrix} \label{sc: GenR}
In real world settings, the assumption of constant between-period correlation may be overly restrictive, as correlation is likely to decay across time. To allow for more realistic time-dependent structures, we generalize the correlation structure of the time-varying cluster-period random effects using a correlation matrix $R$. We propose extending the cluster-period random effect $\nu_{ij}$ into a vector for cluster $i$ denoted as
\begin{equation} \label{eq:nu}
    \boldsymbol{\nu}_i = (\nu_{i1}, \nu_{i2}, \ldots, \nu_{iT})^T \sim N(0, \sigma^2_\nu R)
\end{equation}
where $R=[r_{jm}]$ is a correlation matrix such that $r_{jm} = \text{Corr}(\nu_{ij}, \nu_{im})$ for all $j,m$. Thus $\text{Cov}(\nu_{ij}, \nu_{im}) = \sigma^2_\nu r_{jm}$ with $r_{jj} = 1$ for all $j$ and $r_{jm} \in [-1,1]$ for all $j,m$, following standard assumptions of correlation matrices. The full individual level outcome model can then be represented as in equation \eqref{eq:fullestsimp} where $\nu_{ij}$ is the time-varying cluster-period effect, structured through $R$ as defined in equation \eqref{eq:nu} and all other terms are as defined previously. The total marginal variance remains as defined in equation \eqref{eq:sigma2}.

\subsubsection{Intra-Cluster Correlations Under $R$} 

Under the flexible $R$ correlation structure, the covariance between two individuals in the same cluster, observed at time points $j$ and $m$ is
\begin{equation*}
    \text{Cov}(Y_{ijk}, Y_{iml}) = \sigma^2_\alpha + \sigma^2_\nu r_{jm}
\end{equation*}
for $k \neq l$. Thus
\begin{equation} \label{eq:rjm}
    \rho_{jm} = \frac{\sigma^2_\alpha + \sigma^2_\nu r_{jm}}{\sigma^2_y},
\end{equation}
holds for all $j$ and $m$. When $j=m, \rho_{jm}$ reduces to $\rho_w$ and when $j\neq m$, $\rho_{jm}$ provides a time-indexed generalization of the between-period correlation.

\subsubsection{Cluster-Period Correlation and Covariance Structure Under $R$} \label{sect:cpc}

The total variance of the cluster-period mean is shown in equation \eqref{eq:eta}. We extend the cluster-period mean covariance to account for the non-zero $R$ off-diagonal contribution of $r_{jm}$ such that
\begin{equation*}
    Cov(\bar{Y}_{ij}, \bar{Y}_{im}) = \sigma^2_\alpha + \sigma^2_\nu r_{jm} + \sigma^2_\zeta
\end{equation*}
and the generalized cluster-period correlation such that
\begin{equation}\label{eq:omega}
    Corr(\bar{Y}_{ij}, \bar{Y}_{im}) = \omega_{jm} = \frac{\sigma^2_\alpha + \sigma^2_\nu r_{jm} + \sigma^2_\zeta}{\eta}
\end{equation}
Under the exchangeable model, where $\nu_{ij}$ is independent across time, $R=I_T$ and $r_{jm}=0$ simplifying to
\begin{equation*}
\omega_{jm} = \omega = \frac{\sigma^2_\alpha + \sigma^2_\zeta}{\eta}    
\end{equation*} 
as defined in equation~\eqref{eq:omegaex}. In this exchangeable structure, $\nu_{ij}$ contributes only to the diagonal entries of the covariance matrix. With a general $R$ matrix, the off-diagonal structure can be non-zero representing the structure of the temporal correlation of the cluster-period effect.

We extend this structure into a full covariance matrix for the cluster-period means of a given cluster $i$
\begin{equation}\label{eq:aim1V}
    Cov(\bar{Y_i})= V = \sigma^2_\nu R + \sigma^2_c I_T + (\sigma^2_\alpha + \sigma^2_\zeta)J_T 
\end{equation}
noting that under the special case of $R=I_T$ under an exchangeable correlation structure, this format is identical to equation~\eqref{eq:Vex}. 

We can also parametrize the covariance matrix in terms of total cluster-period variance and correlation such that for $V=[v_{jm}]$:
\begin{equation*}
v_{jm} =\begin{cases} \eta & j = m \\
                    \eta \cdot \omega_{jm} &  j \neq m
       \end{cases} .
\end{equation*}
Alternatively, for $\Omega = [\omega_{jm}]$ we can find a parallel structure to equation~\eqref{eq:Vexeta}
\begin{equation}
    V = \eta(I_T+(1-I_T)\circ \Omega) .
\end{equation}
To provide context to these general formulations, we consider common choices of $R$ that arise in stepped wedge designs and examine their implications for cluster-period dependence and covariance estimation.

\begin{figure}
    \centering
    \includegraphics[width=0.8\linewidth]{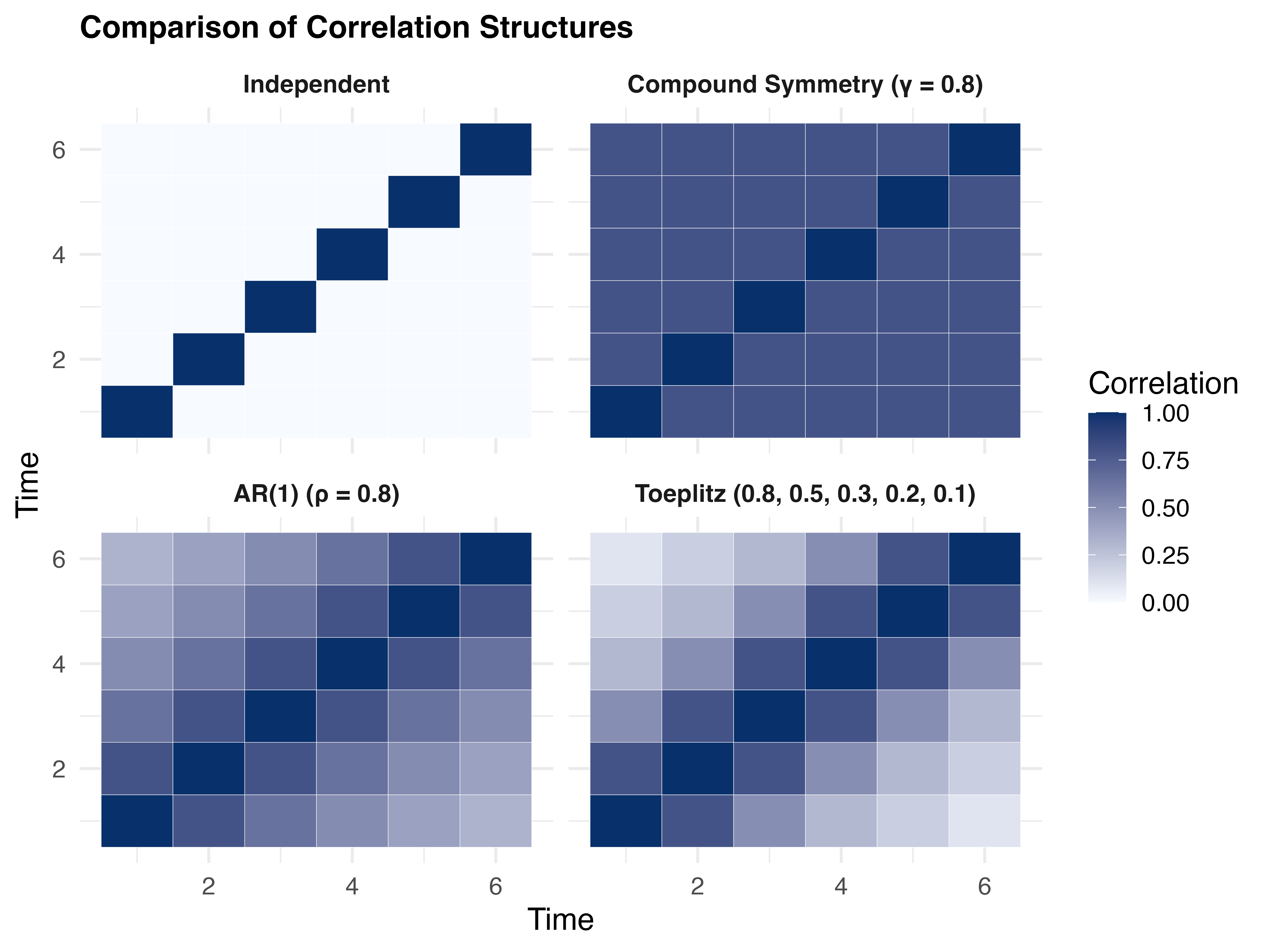}
    \caption[Correlation Structure Heatmaps]{Heatmaps of four correlation structures (Independence, Compound Symmetry, AR(1), and Toeplitz) for T=6 time periods. Lighter colors indicate weaker correlation and darker colors indicate stronger correlation, reaching a maximum of 1 for the within cluster-period correlation on the diagonals.}
    \label{fig:corr_1}
\end{figure}

\subsection{Common Correlation Structures for the $R$ Matrix in Stepped Wedge Designs} \label{sect:commoncor}
Given a defined $r_{jm}$, the individual-level ICC $\rho_{jm}$ and cluster-period mean correlation $\omega_{jm}$ follow directly from equations \eqref{eq:rjm} and \eqref{eq:omega}. Similarly, all covariance results follow from equation \eqref{eq:aim1V} by substitution of the specified correlation matrix $R$ and thus in the following subsection we focus on $R$ and its implications.

\subsubsection{Exchangeable Correlation Structures}
\paragraph{Independent}
When the cluster-period effects, $\nu_{ij}$, are independent across time, $R=I_T$ with $r_{jm}=0$ for $j \neq m$ and $\text{Cov}(\nu_{ij}, \nu_{jm})=0$ for $j \neq m$, is as shown in Figure \ref{fig:corr_1}.This yields the standard two-ICC model described in Section \ref{sec:genICC} and implies that there is no temporal correlation beyond the persistent cluster effect $\alpha_i$, and no off-diagonal contribution from $\nu_{ij}$ to the cluster-period covariance matrix. 

\paragraph{Compound Symmetry}
A compound symmetry (CS) structure assumes that the time-varying cluster-period effects $\nu_{ij}$ are equally correlated across all time points, as shown in Figure \ref{fig:corr_1}. The corresponding $R$ matrix can be written as
\[ R = R_{\text{CS}} = \begin{pmatrix}
    1 & \gamma & \gamma & \ldots & \gamma \\
    \gamma & 1 & \gamma & \ldots & \gamma \\
    \vdots & \vdots & \vdots & \ddots &\vdots \\
    \gamma & \gamma &\gamma & \ldots & 1
\end{pmatrix}
= (1-\gamma)I_T + \gamma J_T
\]
where $\gamma \in [0,1)$ is the constant pairwise correlation between time-varying cluster-period effects, $I_T$ is the $T \times T$ identity matrix, and $J_T$ is the $T \times T$ matrix of ones. Thus, $r_{jm}=1$ for all $j=m$ and $r_{jm}=\gamma$ when $j \neq m$ implying that all time points are equally correlated regardless of temporal distance.  

Under compound symmetry, the correlation between individual-level outcomes in the same cluster, the ICC, is
\begin{equation*}
    \rho_{jm} = \frac{\sigma^2_\alpha + \sigma^2_\nu \gamma}{\sigma^2_y}
\end{equation*}
for all $j, m$ and the correlation between cluster-period means is
\begin{equation*}
    \omega_{jm} = \frac{\sigma^2_\alpha + \sigma^2_\nu \gamma + \sigma^2_\zeta}{\eta}
\end{equation*}
for $j \neq m$. 
The covariance matrix of the cluster-period means can then be derived through the substitution of $R_{\text{CS}}$ into equation \eqref{eq:aim1V}. Due to the constant off-diagonal structure of $R$, the covariance matrix $V$ exhibits uniform pairwise correlation across all time points. Compound symmetry generalizes the independent correlation structure and is commonly used due to its simplicity and closed-form properties \cite{li_mixed-effects_2021, Hu_NR_CS_2018, sundin_power_2022}. 

\subsubsection{Distance-Dependent Correlation Structures}
In many stepped wedge designs, correlation between cluster-period effects is expected to decay with time, motivating the consideration of distance-dependent structures. 
\paragraph{AR(1)}

An autoregressive structure of order 1 (AR(1)) assumes that the correlation between time-varying cluster-period effects decays exponentially with the temporal distance between periods, as shown in Figure \ref{fig:corr_1}. The corresponding correlation matrix, $R$, is
\[
R = R_{\text{AR(1)}} = \begin{pmatrix}
1 & \phi & \phi^2 & \ldots & \phi^{T-1} \\
\phi & 1 & \phi & \ldots & \phi^{T-2} \\
\phi^2 & \phi & 1 & \ldots & \phi^{T-3} \\
\vdots & \vdots & \vdots & \ddots & \vdots \\
\phi^{T-1} & \phi^{T-2} & \phi^{T-3} & \ldots & 1
\end{pmatrix}
\]
where $\phi \in (0,1)$ determines the strength and rate of decay of the temporal correlation. Element-wise, this corresponds to $r_{jm} = \phi^{|j-m|}$. Periods that are closer in time are more strongly correlated than those farther apart. Substituting  $R_{\text{AR(1)}}$ into equation \eqref{eq:aim1V} induces a distance-dependent cluster-period covariance structure for AR(1) structures. The AR(1) structure offers more realistic modeling of temporal dependence than the compound symmetry or independent structure by allowing correlations to decrease as time between periods increases.

\subsubsection{Proportional Decay}
The proportional decay structure generalizes the AR(1) structure by introducing an additional scaling parameter, $\lambda$, allowing for weaker immediate correlation between adjacent time periods. The correlation matrix, $R$, for the time-varying cluster-period effect under the proportional decay correlation structure is
\[
R = R_\text{PD}=
\begin{pmatrix}
1 & \lambda\phi & \lambda\phi^2 & \ldots & \lambda\phi^{T-1} \\
\lambda\phi & 1 & \lambda\phi & \ldots & \lambda\phi^{T-2} \\
\lambda\phi^2 & \lambda\phi & 1 & \ldots & \lambda\phi^{T-3} \\
\vdots & \vdots & \vdots & \ddots & \vdots \\
\lambda\phi^{T-1} & \lambda\phi^{T-2} & \lambda\phi^{T-3} & \ldots & 1
\end{pmatrix}
\]
where $\lambda \in (0,1)$ is the scaling parameter and $\phi \in (0,1)$ determines the strength and rate of decay of the temporal correlation. Element-wise, this corresponds to $r_{jm}=\lambda \cdot \phi^{|j-m|}$. Substituting  $R_{\text{PD}}$ into equation \eqref{eq:aim1V} induces a distance-dependent cluster-period covariance structure for proportional decay structures. By relaxing the assumption of perfect immediate correlation, the proportional decay model offers a more conservative alternative to AR(1) while still allowing for a gradual reduction of correlation over time. 

\subsubsection{Toeplitz}
The Toeplitz structure is a general, symmetric correlation structure that defines correlations between time-varying cluster-period effects based on the temporal lag between periods, with a distinct parameter for each lag, as shown in Figure \ref{fig:corr_1}. The corresponding correlation matrix, $R$, is
\[
R = R_{\text{Tz}} = 
\begin{pmatrix}
    1 & \delta_1 & \delta_2 & \ldots & \delta_{T-1} \\
    \delta_1 & 1 &\delta_1 & \ldots & \delta_{T-2} \\
    \delta_2 & \delta_1 & 1 & \ldots & \delta_{T-3} \\
    \vdots & \vdots & \vdots & \ddots & \vdots \\
    \delta_{T-1} & \delta_{T-2} & \delta_{T-3} & \ldots & 1
\end{pmatrix}
\]
such that $r_{jm} = \delta_h$ with $h=|j-m|$ is the lag. We assume $\delta_0 = 1$ for identifiability, corresponding to perfect within-period correlation. Under the Toeplitz structure, all time pairs that are the same distance apart share the same correlation value, but these values are not constrained to a specific decay pattern such as AR(1) or proportional decay. Substituting  $R_{\text{Tz}}$ into equation \eqref{eq:aim1V} induces a distance-dependent cluster-period covariance structure for Toeplitz structures. This structure is highly flexible and can accommodate irregular or non-monotonic patterns of correlation across time. The Toeplitz structure is useful when there is a more complex correlation structure than can be captured by AR(1) or proportional decay. However, Toeplitz models require the estimation of more parameters and may be less reliable in small sample size settings \cite{YANG_2021_toep}.

\subsection{Implications for Estimating Treatment Effects in Multiple-Intervention SWDs}\label{sect:implcov}

Because the variance of the treatment effect estimators depends on the inverse of the cluster-period covariance matrix, $V$, assumptions about $R$ directly determine efficiency and power in multiple-intervention stepped wedge designs. Table~\ref{tab:var_tab} presents key variance and correlation expressions under common correlation structures for the time-varying cluster-period effect $R$ in SWDs. The specification of $R$ directly determines the cluster-period covariance structure, and, consequently the variance of the treatment effect estimators, $\hat{\theta}$. In Section \S\ref{sec:sim} we use simulations and analytic results to evaluate the performance of variance estimators under these structures, and the consequences of misspecifying $R$ in the context of multiple-intervention stepped wedge designs. 

\begin{table}[!ht]
\caption[Key variance and correlation quantities by correlation structure]{Key variance and correlation quantities by correlation structure.
$^{\dagger}$ Number of free parameters required to define correlation matrix $R$.
$^{\ddagger}$ Whether closed-form expressions for $\mathrm{Var}(\hat{\theta})$ are available analytically.
All entries for $r_{jm}$, $\rho_{jm}$, and $\omega_{jm}$ are for $j\neq m$; for $j=m$, $r_{jj}=1$ and $\rho_{jj}=\rho_w$.}
\label{tab:var_tab}
\centering
\small
\setlength{\tabcolsep}{5pt}
\renewcommand{\arraystretch}{1.5}
\setlength{\extrarowheight}{1.5pt}
\begin{tabular}{lccccc}
\hline
\textbf{Quantity} & \textbf{Exchangeable} & \textbf{CS} & \textbf{AR(1)} & \textbf{PD} & \textbf{Toeplitz} \\
\hline
$r_{jm}$ & $0$ & $\gamma$ & $\phi^{|j-m|}$ & $\lambda\,\phi^{|j-m|}$ & $\delta_{|j-m|}$ \\

$\rho_{jm}$ & $\frac{\sigma_\alpha^2}{\sigma_y^2}$ &
$\frac{\sigma_\alpha^2 + \sigma_\nu^2\gamma}{\sigma_y^2}$ &
$\frac{\sigma_\alpha^2 + \sigma_\nu^2\phi^{|j-m|}}{\sigma_y^2}$ &
$\frac{\sigma_\alpha^2 + \sigma_\nu^2\lambda\phi^{|j-m|}}{\sigma_y^2}$ &
$\frac{\sigma_\alpha^2 + \sigma_\nu^2\delta_{|j-m|}}{\sigma_y^2}$ \\

$\omega_{jm}$ & $\frac{\sigma_\alpha^2 + \sigma_\zeta^2}{\eta}$ &
$\frac{\sigma_\alpha^2 + \sigma_\nu^2\gamma + \sigma_\zeta^2}{\eta}$ &
$\frac{\sigma_\alpha^2 + \sigma_\nu^2\phi^{|j-m|} + \sigma_\zeta^2}{\eta}$ &
$\frac{\sigma_\alpha^2 + \sigma_\nu^2\lambda\phi^{|j-m|} + \sigma_\zeta^2}{\eta}$ &
$\frac{\sigma_\alpha^2 + \sigma_\nu^2\delta_{|j-m|} + \sigma_\zeta^2}{\eta}$ \\

$\mathrm{Cov}(\bar{Y}_i)=V$ &
\multicolumn{5}{c}{$\sigma_\nu^2 R + \sigma_c^2 I_T + (\sigma_\alpha^2+\sigma_\zeta^2)J_T$} \\

$R$ & $I_T$ & $R_{\mathrm{CS}}$ & $R_{\mathrm{AR(1)}}$ & $R_{\mathrm{PD}}$ & $R_{\mathrm{Tz}}$ \\

$\#$ params in $R^{\dagger}$ & $0$ & $1$ & $1$ & $2$ & $T-1$ \\

Closed-form$^{\ddagger}$ & Yes & Yes & Yes & Yes & Yes \\
\hline
\end{tabular}
\end{table}

\subsection{Covariance Estimation and Power Analysis Framework for multiple-intervention SWD with Flexible Correlation Structures}\label{sec:cov229}

\subsubsection{Notation}
We now express these results in matrix form to derive variance and power expressions for treatment effect estimators under flexible correlation structures. We define the outcome vector $\mathbf{Y} = (\bar{Y}_{11}, \ldots, \bar{Y}_{iT}, \ldots , \bar{Y}_{I1}, \ldots, \bar{Y}_{IT})' $. Under the assumption of cluster independence, we represent $\mathbf{V}$, the variance-covariance matrix of $\mathbf{Y}$ as an $IT \times IT$ matrix
\renewcommand{\arraystretch}{0.8}
\[
\mathbf{V} = 
\begin{pmatrix}
   \mathbf{ V}_1 & 0 & 0 & \ldots & 0 \\
    0 & \mathbf{V}_2 & 0 & \ldots & 0 \\
    0 & 0 & \mathbf{V}_3 & \ldots & 0 \\
    \vdots & \vdots & \vdots & \ddots & \vdots \\
    0 & 0 & 0 & \ldots & \mathbf{V}_I
\end{pmatrix}
\]
with each $\mathbf{V}_i$ being the corresponding covariance matrix for cluster-period mean $\bar{Y}_i$ for cluster $i$, 
\begin{equation*}
    \mathbf{V_i} = \sigma^2_\nu R + \sigma^2_cI_T  + (\sigma^2_\alpha + \sigma^2_\zeta)J_T 
\end{equation*}
with $R \in \mathbb{R}^{T\times T}$ and all other terms as defined previously. 

We define the $(T + Q) \times 1$ regression coefficient vector, $\boldsymbol{\Theta}$ for fixed effects as 
\begin{equation*}
\boldsymbol{\Theta} = 
\begin{bmatrix}
    \mu & \beta_1 & \ldots & \beta_{T-1} & \theta_1 & \ldots & \theta_Q 
\end{bmatrix}'
\end{equation*}
where we set $\beta_T = 0$ for identifiability. Under the common case of two interventions with an interaction compared to one control, we can write this vector as
\begin{equation*}
\boldsymbol{\Theta} = 
\begin{bmatrix}
    \mu & \beta_1 & \ldots & \beta_{T-1} & \theta_1 & \theta_2 & \theta_3 
\end{bmatrix}'. 
\end{equation*}

To generalize the treatment effect estimation to multiple-intervention stepped wedge designs, we extend the classic linear-mixed model precision matrix formulation $\mathbf{C}=\mathbf{Z}'\mathbf{V}^{-1}\mathbf{Z}$ to accommodate non-exchangeable correlation structures as follows. 

 The fixed effects design matrix $\mathbf{Z} \in \mathbb{R}^{IT} \times (T + Q)$ is
\[
\mathbf{Z} = 
\begin{bmatrix}
    \mathbf{Z}_1 \\
    \mathbf{Z}_2 \\
    \vdots \\
    \mathbf{Z}_I
\end{bmatrix}
\]
where each $\mathbf{Z}_i \in T \times (T+Q)$ is of the form
\renewcommand{\arraystretch}{0.4}
\[
\mathbf{Z}_i = 
\begin{bmatrix}
    & \mathbf{I}_{T-1} & & &  \\   \mathbf{1}_T && \mathbf{X}_{1i} & \ldots & \mathbf{X}_{Qi} \\
   & \mathbf{0}'_{T-1} & & &
\end{bmatrix}
\] 
and $\mathbf{X}_{qi} = (X_{qi1}, X_{qi2}, \ldots, X_{qiT} )$ is a vector of indicators of whether at time $j$, cluster $i$ received treatment $q$. The matrix $\mathbf{I}_{T-1}$ contains indicators for each time point $j$ from $1$ to $T-1$ corresponding to fixed effects for time and the vector $\mathbf{0}'_{T-1}$ corresponds to time $T$ for identifiability. Under the common case of two interventions with an interaction, we can simplify this to
\[
\mathbf{Z}_i = 
\begin{bmatrix}
    & \mathbf{I}_{T-1} & & &  \\   \mathbf{1}_T && \mathbf{X}_{1i} & \mathbf{X}_{2i} & (\mathbf{X}_1 \mathbf{X}_2)_i \\
   & \mathbf{0}'_{T-1} & & &
\end{bmatrix}.
\] 

Under the exchangeable variance structure, $\mathbf{R}=\mathbf{I}_T$, or the compound symmetry structure, $\mathbf{R}=(1-\gamma)\mathbf{I}_T + \gamma \mathbf{J}_T$, $\mathbf{V}_i$ is symmetric and thus we are able to derive a closed-form inverse, $\mathbf{V}_i^{-1}$ using the Sherman-Morrison identity. Sundin and Crespi \cite{sundin_power_2022} have shown the resulting structure for exchangeable variance and below we will derive the inverse under compound symmetry. For other $R$ matrices such as AR(1), proportional decay, and Toeplitz structures, we are able to compute closed-form inverses using the Usmani Inversion and Sherman-Morrison rank one update.

\subsubsection{Variance Estimation Under Compound Symmetry Correlation}\label{sec:CSvar}
Under compound symmetry with $\mathbf{R} = (1-\gamma)\mathbf{I}_T+\gamma \mathbf{J}_T$ we can reparameterize the block covariance such that it follows the form $\mathbf{V}_i=a\mathbf{J}_T + b\mathbf{I}_T$:
\begin{align*}
    \mathbf{V}_i &= \sigma^2_\nu[(1-\gamma)\mathbf{I}_T+\gamma \mathbf{J}_T] + (\sigma^2_\alpha + \sigma^2_\zeta)\mathbf{J}_T + \sigma^2_c \mathbf{I}_T \\
    &= (\sigma^2_\alpha + \sigma^2_\zeta + \sigma^2_\nu\gamma)\mathbf{J}_T + (\sigma^2_c + (1-\gamma)\sigma^2_\nu)\mathbf{I}_T \\
    &=a\mathbf{J}_T + b\mathbf{I}_T ,
\end{align*}
with $a = \sigma^2_\alpha + \sigma^2_\zeta + \sigma^2_\nu\gamma$ and $b=\sigma^2_c + (1-\gamma)\sigma^2_\nu $ which allows us to invert $\mathbf{V}_i$ following the Sherman-Morrison inversion. We can thus invert $\mathbf{V}_i$ such that
\begin{align*}
    \mathbf{V}_i^{-1} &= \frac{1}{b}\mathbf{I}_T - \frac{a}{b(b+aT)}\mathbf{J}_T \\
    &= \frac{1}{\sigma^2_c+(1-\gamma)\sigma^2_\nu}\mathbf{I}_T - \frac{\sigma^2_\alpha + \sigma^2_\zeta + \sigma^2_\nu\gamma}{[\sigma^2_c + (1-\gamma)\sigma^2_\nu ][\sigma^2_c + (1-\gamma)\sigma^2_\nu + (\sigma^2_\alpha + \sigma^2_\zeta + \sigma^2_\nu\gamma )T]}\mathbf{J}_T .
\end{align*}
Thus for each cluster $i$, $\mathbf{V}_i^{-1}$ has constant diagonal elements $\frac{1}{b} - \frac{a}{b(b+aT)}$ and constant off-diagonal elements $\frac{-a}{b(b+aT)}$ and the full $\mathbf{V}^{-1}_i$ is block diagonal across clusters. Since $\mathbf{V}^{-1}$ is block diagonal we can write
\begin{equation*}
    \mathbf{Z'V}^{-1}\mathbf{Z} = \sum_{i=1}^I\mathbf{Z}_i'\mathbf{V}_i^{-1}\mathbf{Z}_i
\end{equation*}
with
\begin{align*}
       \mathbf{Z}'_i \mathbf{V}_i^{-1}\mathbf{Z}_i&= \frac{1}{b}\mathbf{Z}_i'\mathbf{Z}_i - \frac{a}{b(b+aT)}\mathbf{Z}_i'\mathbf{J}_T \mathbf{Z}_i \\
    &= \frac{1}{\sigma^2_c+(1-\gamma)\sigma^2_\nu}\mathbf{Z}_i'\mathbf{Z}_i  - \frac{\sigma^2_\alpha + \sigma^2_\zeta + \sigma^2_\nu\gamma}{[\sigma^2_c + (1-\gamma)\sigma^2_\nu] [\sigma^2_c + (1-\gamma)\sigma^2_\nu + (\sigma^2_\alpha + \sigma^2_\zeta + \sigma^2_\nu\gamma )T]}\mathbf{Z}_i'\mathbf{J}_T \mathbf{Z}_i .
\end{align*}
For a given design matrix $\mathbf{Z_i}$ we can invert the block matrix to solve for the corresponding regression coefficient,
\begin{equation*}
    \text{Cov}(\hat{\boldsymbol{\Theta}}) =\Big[ \sum_{i=1}^I [\frac{1}{b}\mathbf{Z}_i'\mathbf{Z}_i - \frac{a}{b(b+aT)}\mathbf{Z}_i'\mathbf{J}_T \mathbf{Z}_i ] \Big]^{-1}.
\end{equation*}

We derive a closed form variance estimator under AR(1) in Section 1 of the Supporting Information.

In the next section, we simulate multiple-intervention stepped wedge designs under varying cluster-period correlations, $R$ to evaluate variance and power across these conditions, as well as to assess the impact of mis-specifying the underlying correlation structure.

\section{Simulation and Analytic Study} \label{sec:sim}

\subsection{Setting}\label{sect:Aim1ResSetting}
In this simulation and analytic study, we investigate the characteristics of the $R$ correlation structures described in Section \S\ref{sect:commoncor}. For compound symmetry and AR(1) structures we present results following the analytic frameworks proposed in sections \S\ref{sec:CSvar} and Section 1 of the Supporting Information, where feasible, and utilized Markov chain Monte Carlo (MCMC) simulations otherwise. Under simulations, all datasets are simulated 10,000 times. We focus on repeated cross-sectional designs ($\sigma^2_\psi=0$) with equal cluster sizes and one cluster per sequence. We assume 50 individuals per cluster-period ($n=50$), ICC parameters $\rho_w=0.2$ and $\rho_b=0.05$, total variance of 2.0 ($\sigma^2_y=2.0$), true Cohen's d standardized main effect value of 0.6 ($\theta_{1,\text{stand}}=\theta_{2,\text{stand}}=0.6$), and true standardized interaction effect value of 0.3 ($\theta_{3,\text{stand}}=0.3$), unless otherwise specified.  We utilized standardized effect sizes using Cohen's d, where Cohen's d is defined as $\frac{\theta}{\sqrt{\sigma^2_y}}$  \cite{cohen1988statistical}. Bonferroni corrections for multiple comparisons were utilized throughout this work. 

\begin{figure}
        \centering
        \includegraphics[width=0.5\linewidth]{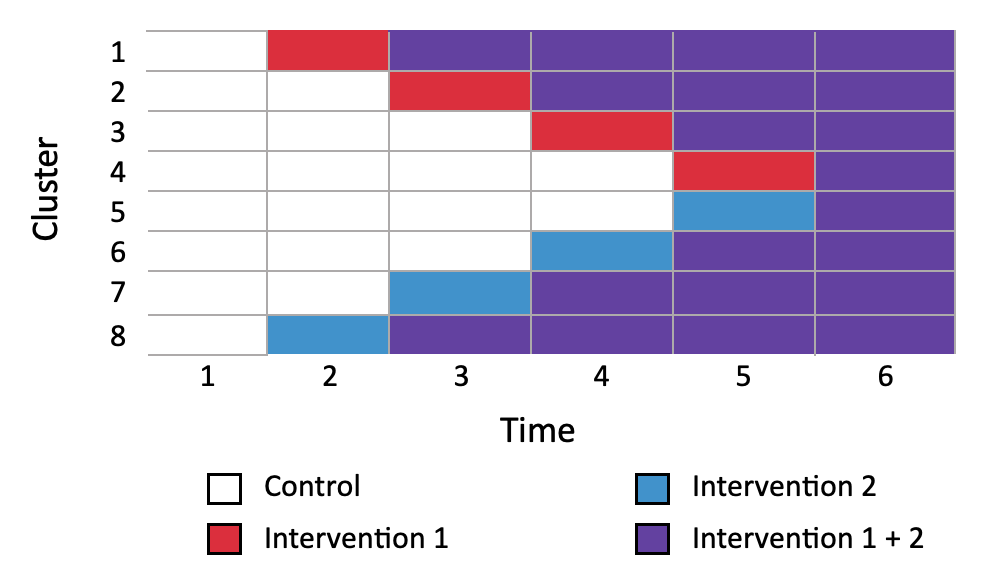}
        \caption[8 Cluster Factorial Design]{Factorial design with 8 cluster and 6 periods utilized for subsequent analytic and simulation studies. White cluster-periods represent control periods, red is intervention 1 only, blue is intervention 2 only, and purple is the interaction for both intervention 1 and intervention 2.}
        \label{fig:aim1res_fact}
    \end{figure}

Results will focus on factorial design M-SWD studies. Unless otherwise specified, factorial designs will include 8 clusters and 6 time points. The standard factorial design that was utilized throughout the following analytic and simulation studies is shown in Figure \ref{fig:aim1res_fact}. 

\subsubsection{Data Generating Model}
Where simulations are utilized, we utilized the individual level data generation model with structure
\begin{equation*}
    Y_{ijk}=\mu +\beta_j + \alpha_i + \nu_{ij} + \psi_{ik} +  \theta_1 X_{1ij} +\theta_2X_{2ij} + \theta_3X_{1ij}X_{2ij}+e_{ijk}
\end{equation*}
with $\alpha_i \sim N(0,\sigma^2_\alpha)$, $\psi_{ik} \sim N(0,\sigma^2_\psi)$, $\nu_{ij}\sim N(0,\sigma^2_\nu R)$, and $e_{ijk} \sim N(0,\sigma^2_e)$ where $\sigma^2_\psi=0$ for repeated cross-sectional designs. This structure represents a standard linear mixed-effects model with a continuous outcome and identity link.

\subsubsection{Analysis Methods}\label{sec:aim1analysismeth}

Details about the analysis methods utilized for this study can be found in Section 2 of the Supporting Information. The approaches are also summarized in Table \ref{tab:aim1analysismeth}.

\begin{table}[ht]
    \centering
        \caption[Analysis Methods]{Summary of analysis methods used in the following sections. $^*$ Ind=Independence; $^{**}$ GLS=Generalized Least Squares ; $^\text{\textdagger}$ C-P = Cluster-Period }
    \label{tab:aim1analysismeth}
    \begin{tabular}{c c c c c c}
        \textbf{Method} & \textbf{Simulation?} & \textbf{True }$\mathbf{V}$? & \textbf{Unit of Analysis} & \textbf{Intended Role}  \\ 
         \hline \\
         Analytical & No & Yes & C-P$^\text{\textdagger}$ Means & Theoretical Benchmark \\ \\
         GLS$^{**}$-Oracle & Yes & Yes & C-P$^\text{\textdagger}$ Means & Empirical Validation \\ \\
         LMM-CS & Yes & No - Estimated (CS) & Individual & Realistic \\ \\
         LMM-Ind$^*$ & Yes & No - Estimated (Ind$^*$) & Individual & Misspecified
         
    \end{tabular}

\end{table}

\section{Simulation and Analytic Study Results} \label{sec:aim1results}
\subsection{Validation of Asymptotic Power Formulas}\label{sec:aim1analy}

We validated the asymptotic variance and standard error expressions derived in Section \S\ref{sec:CSvar} using simulations, finding very similar results between simulation and theoretical power (see Section 3 in the Supporting Information). 

\subsection{Power Under Correct Correlation Specification}\label{sec:aim1res1}

\begin{figure}
    \centering
    \includegraphics[width=1\linewidth]{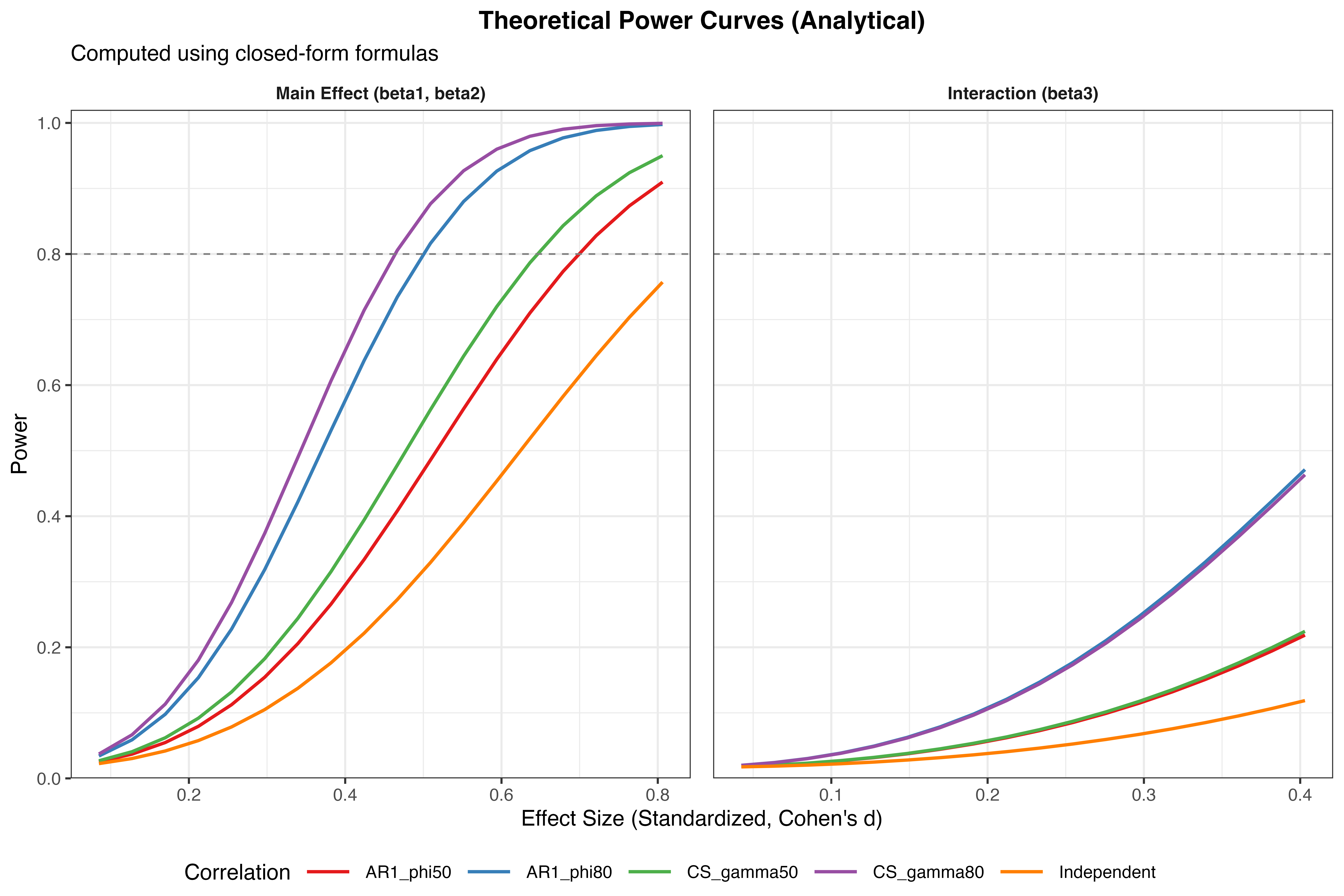}
    \caption[Analytic Power Curves Across Correlation Structures]{Analytic power curves as a function of standardized effect size for main and interaction effects under correct specification of the cluster-period correlation structure. Results for an 8 cluster, 6 period, symmetric factorial repeated-cross section design with $n=50$ individuals per cluster-period, $\rho_b=0.05$ and $\rho_w=0.2$. Curves are computed using closed-form variance expressions, with colors indicating the assumed (and true) $R$ correlation structure. The dashed horizontal line at 0.8 denotes the conventional 80\% power target. Main effects are shown in a single panel due to symmetry between main effect 1 and main effect 2, which yield identical theoretical power curves.}
    \label{fig:Aim1Q1_fig2}
\end{figure}

Figure \ref{fig:Aim1Q1_fig2} presents analytic power curves for main and interaction effects under independence, compound symmetry, and AR(1) correlation structures, computed using the closed-form variance expressions derived in Section \S\ref{sec:CSvar} and Section 1 in the Supporting Information. All results correspond to the same underlying design and variance components, as described in Section \S\ref{sect:Aim1ResSetting}, such that differences arise solely from the assumed correlation structure. Power to detect the interaction effect was uniformly lower than the main effects across all correlation structures. The interaction power exhibited heightened sensitivity to the assumed correlation structure, with larger relative differences across structures than were observed for the main effects. 

\begin{figure}
    \centering
    \includegraphics[width=1\linewidth]{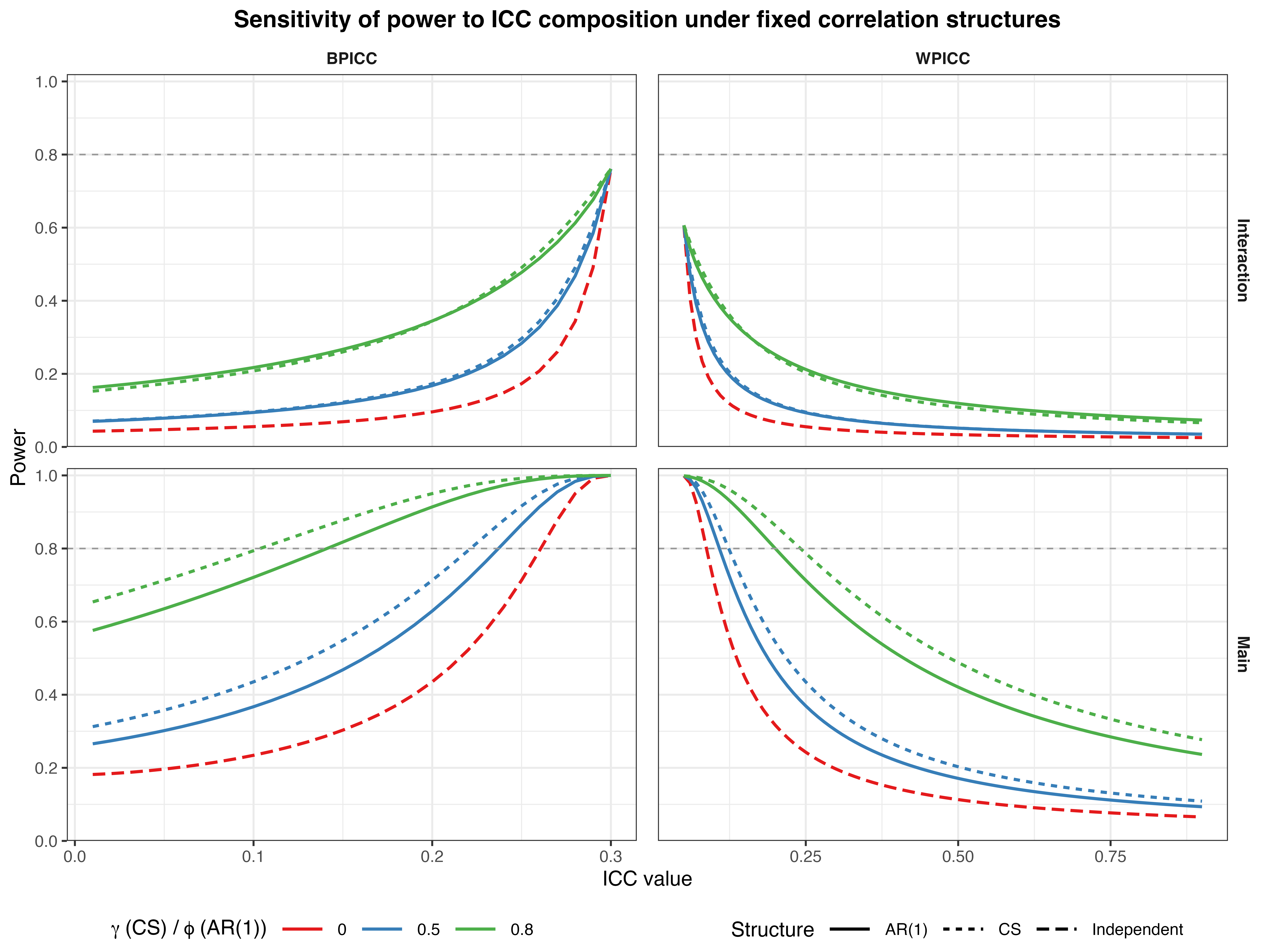}
    \caption[Sensitivity of Power to ICC Composition]{Sensitivity of theoretical power to ICC composition under fixed correlation structures. Power is shown as a function of the between-period intracluster correlation (BPICC, $\rho_b$, left column) and the within-period intracluster correlation (WPICC, $\rho_w$, right column), holding all other design parameters fixed. When WPICC varies, BPICC is held constant at 0.05. When BPICC varies, WPICC is held constant at 0.3. Results are shown separately for main effects (bottom row) and interaction effects (top row) with main effects computed at a standardized effect size 0.5 and interaction effects at 0.3. Results for an 8 cluster, 6 period, symmetric factorial repeated-cross section design with $n=50$ individuals per cluster-period and total variance $\sigma^2_y=2$. Colors indicate the magnitude of the correlation parameter ($\gamma$ for compound symmetry and $\phi$ for AR(1)). The dashed horizontal line denotes the conventional 80\% power threshold.}
    \label{fig:Aim1Q1ICC}
\end{figure}

To clarify the mechanism underlying the sensitivity of power to the assumed correlation structure observed in previous sections, Figure \ref{fig:Aim1Q1ICC} examines how power varies with the composition of the intracluster correlation, holding the $R$ correlation structures and effect sizes fixed. Across independence, compound symmetry, and AR(1) structures, power was substantially more sensitive to changes in the within-period ICC ($\rho_w$) than to changes in the between-period ICC ($\rho_b$). For main effects, increasing $\rho_w$ led to notable gains in power and emphasizes the differences between correlation structures and parameter values, whereas variations in $\rho_b$ produced comparatively modest changes. This reflects the fact that only the cluster-period component of variability ($\sigma^2_\nu$) directly interacts with the temporal correlation structure $R$.

To assess whether the previous results regarding cluster-period correlation structure extend beyond the repeated cross-sectional designs considered above, we examined a cohort design in which the same individuals were followed longitudinally across time. We quantified the impact of individual-level correlation across time through the individual autocorrelation (IAC) where the IAC is the proportion of individual-level variance attributable to a time-invariant individual effect.  The IAC can be represented as 
\begin{equation*}
    \text{IAC} = \pi = \frac{\sigma^2_\psi}{\sigma^2_\psi + \sigma^2_e}
\end{equation*} 
where $\sigma^2_\psi$ is the variance of the individual-level random effect and $\sigma^2_e$ is the individual-level residual variance. For repeated cross-sectional multiple-intervention SWDs, as examined throughout the rest of this work, $\pi=0$.

\begin{figure}
    \centering
    \includegraphics[width=1\linewidth]{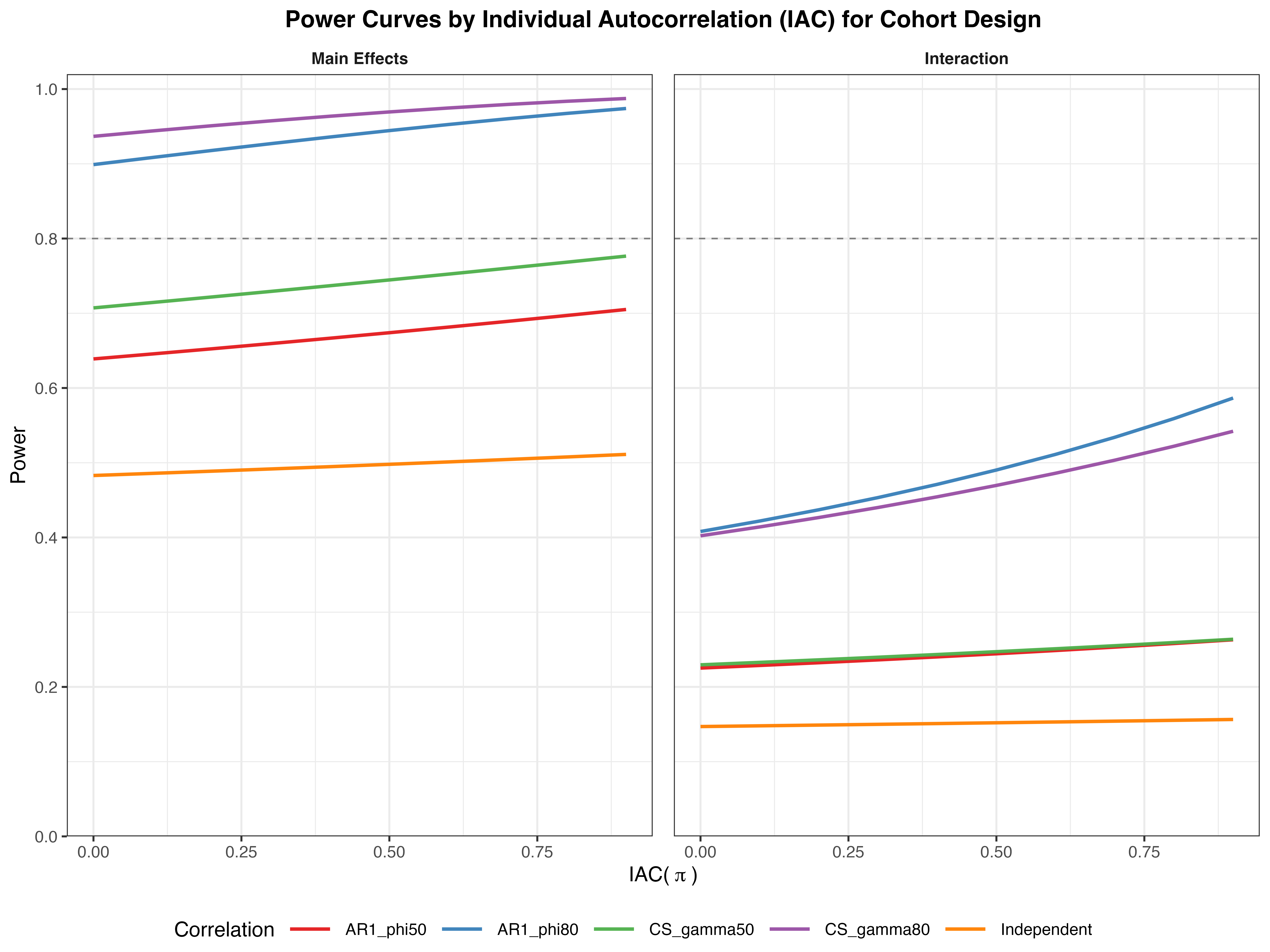}
    \caption[Power Curves by Individual Autocorrelation for Cohort Design]{Theoretical power as a function of individual autocorrelation ($\pi$) under a cohort design for main effects (left) and interaction effect (right). Power is shown for fixed standardized effect sizes (0.6 for main; 0.3 for interaction) under independence, compound symmetry, and AR(1) cluster-period correlation structures. Results are computed using closed-form variance expressions under correct specification of the correlation structure $R$. Results are shown for an 8 cluster, 6 period symmetric factorial design with $n=50$ individuals per cluster-period, $\rho_b=0.05$, and $\rho_w=0.2$. The dashed horizontal line indicates the conventional 80\% power threshold. $\pi=0$ corresponds to a repeated cross-sectional design.}
    \label{fig:Aim1_Q1_pi}
\end{figure}

Figure \ref{fig:Aim1_Q1_pi} shows theoretical power as a function of $\pi$ for fixed effect sizes under independence, compound symmetry, and AR(1) correlation structures. For both main and interaction effects, power increased monotonically, but modestly, with $\pi$, reflecting the efficiency gains acquired from repeated measurements on the same individuals. Consequently, the conclusions drawn throughout this work under repeated cross-sectional designs regarding the impact of $R$ extend naturally to cohort settings. 

\subsection{Robustness of Treatment Effect Estimation to Correlation Misspecification}\label{sec:aim1res2}

This section examines the robustness of treatment effect estimation and inference in M-SWDs when the working correlation structure differs from the true cluster-period correlation. Across all scenarios, the mean model, and thus the true treatment effects and fixed time effects, were correctly specified and differences arise solely from misspecification of the analysis covariance structure. Details about the analysis methods utilized can be found in Section 2 of the Supporting Information.

\begin{figure}
    \centering
    \includegraphics[width=1\linewidth]{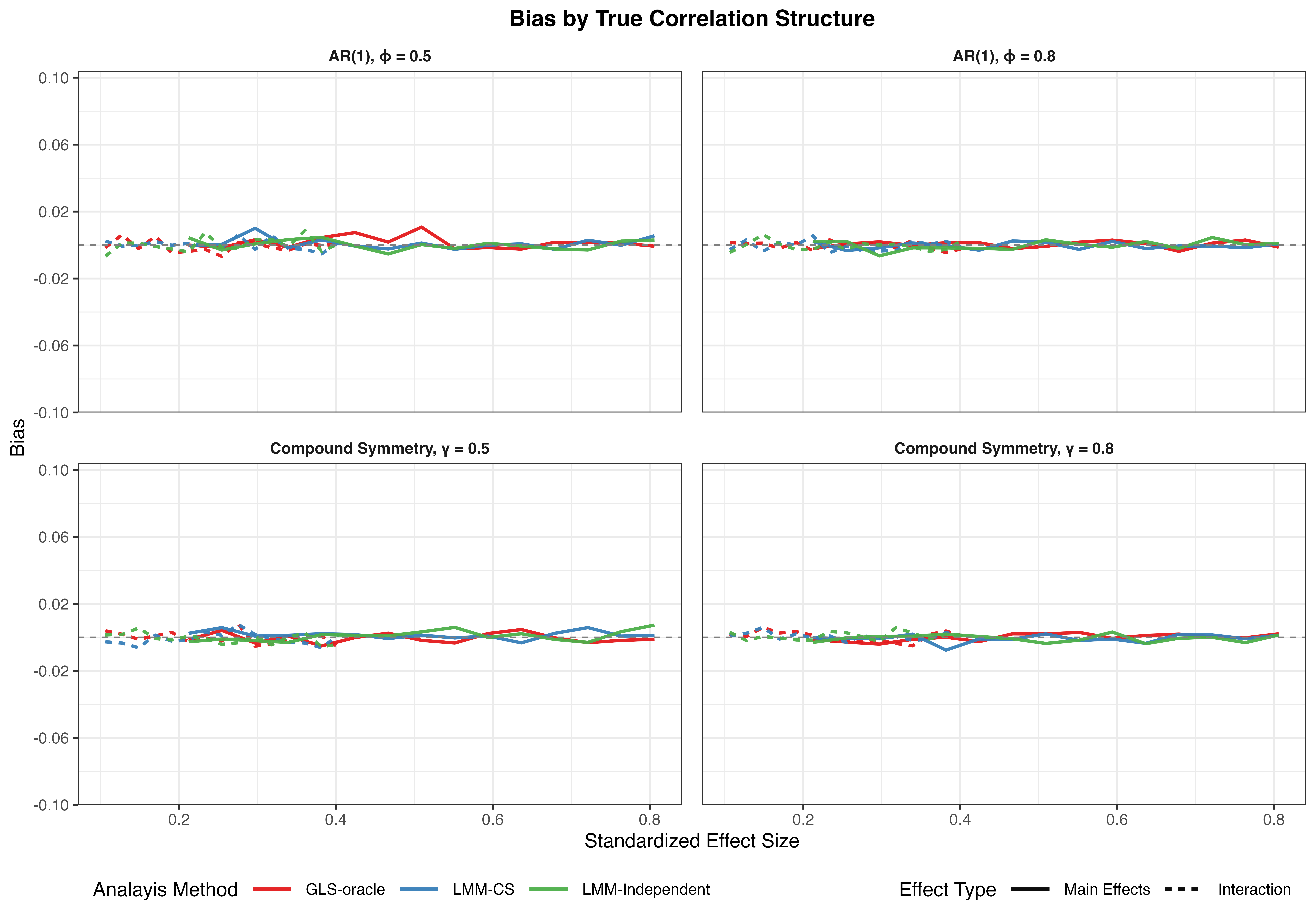}
    \caption[Bias Under Correlation Misspecification]{Empirical bias of fixed-effect estimators across standardized effect sizes, stratified by the true cluster-period correlation structure ($R$). Results are shown for main effects and interaction effects under three simulation-based analysis methods: GLS-oracle, LMM-CS, and LMM-independent. Panels correspond to AR(1) correlation with autocorrelation parameter $\phi\in\{0.5,0.8\}$ and compound symmetry with correlation parameter $\gamma \in \{0.5,0.8\}$. The horizontal dashed line denotes 0 bias. Results for an 8 cluster, 6 period, symmetric factorial repeated-cross section design with $n=50$ individuals per cluster-period, $\rho_b=0.05$ and $\rho_w=0.2$.}
    \label{fig:aim1biasmis}
\end{figure}

Figure \ref{fig:aim1biasmis} displays the empirical bias of fixed-effect estimators across a range of standardized effect sizes under each true correlation structure. Across all scenarios, bias was negligible (max absolute bias $<0.02$) for both main and interaction effects, regardless of the analysis method used.

\begin{figure}
    \centering
    \includegraphics[width=\linewidth]{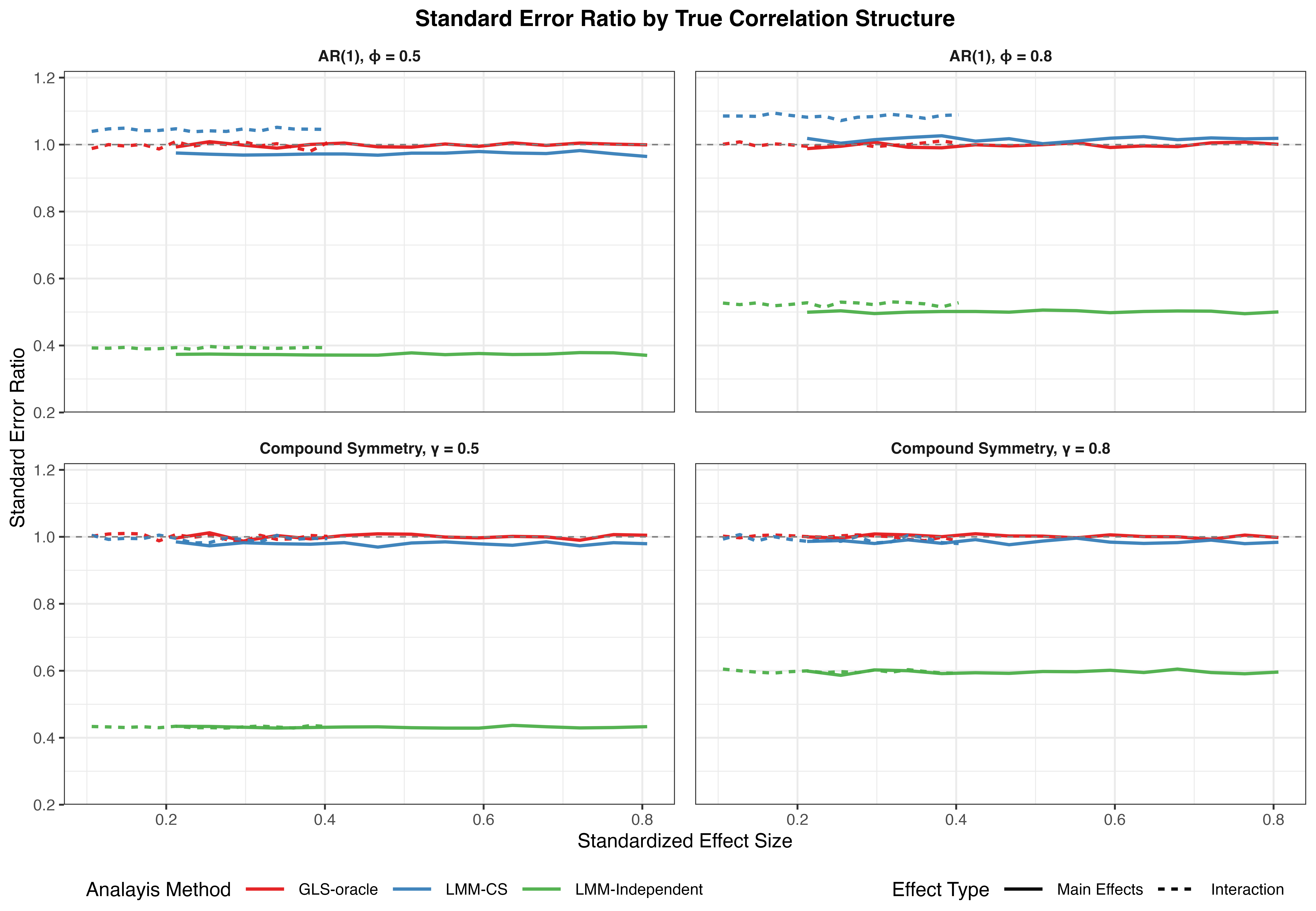}
    \caption[Standard Error Ratio Under Correlation Misspecification]{Ratio of estimated standard errors to true standard errors across standardized effect sizes, stratified by the true cluster-period correlation structure. Results are shown for main effects (solid lines) and interaction effects (dashed lines) under three simulation-based analysis methods: GLS-Oracle, LMM-CS, and LMM-Independent. Panels correspond to AR(1) correlation with autocorrelation parameter $\phi\in\{0.5,0.8\}$ and compound symmetry with correlation parameter $\gamma \in \{0.5,0.8\}$. The horizontal reference line denotes a ratio of 1. Results for an 8 cluster, 6 period, symmetric factorial repeated-cross section design with $n=50$ individuals per cluster-period, $\rho_b=0.05$ and $\rho_w=0.2$.}
    \label{fig:aimseratiomis}
\end{figure}

Figure \ref{fig:aimseratiomis} presents the ratio of estimated standard errors to true standard errors across a range of standardized effect sizes. For both AR(1) and compound symmetry correlation structures, the GLS-oracle and LMM-CS approaches yielded standard errors that were close to the true benchmark (SE ratio range: $0.96 - 1.03$). In contrast, the LMM-Independent approach exhibited substantial underestimation of standard errors when the true cluster-period correlation was non-zero (SE ratio range: $0.37 - 0.61$). Within each correlation structure, the standard error ratio under the independence working model increased with the correlation parameter $\gamma$ or $\phi$, but remained far below both the oracle and LMM-CS methods, even under strong correlation, indicating persistent underestimation of variance.

\subsection{Design-Stage Consequences of Correlation Misspecification}\label{sec:aim1designimp}

The results in Section \S\ref{sec:aim1res2} focus on analysis-stage misspecification, demonstrating that incorrect modeling of the cluster-period correlation structure can substantially distort variance estimation and inference, even when the mean model is correctly specified. In practice, however, assumptions about correlation structure are also important in the design stage, where power and sample size calculations are often conducted under simplifying assumptions such as independence. In this subsection, we examine the consequences of design-stage correlation misspecification, holding the stepped wedge design fixed and isolating the impact of planning under an incorrect correlation structure when the true data-generating process exhibits cluster-period correlation. 

\begin{figure}
    \centering
    \includegraphics[width=0.8\linewidth]{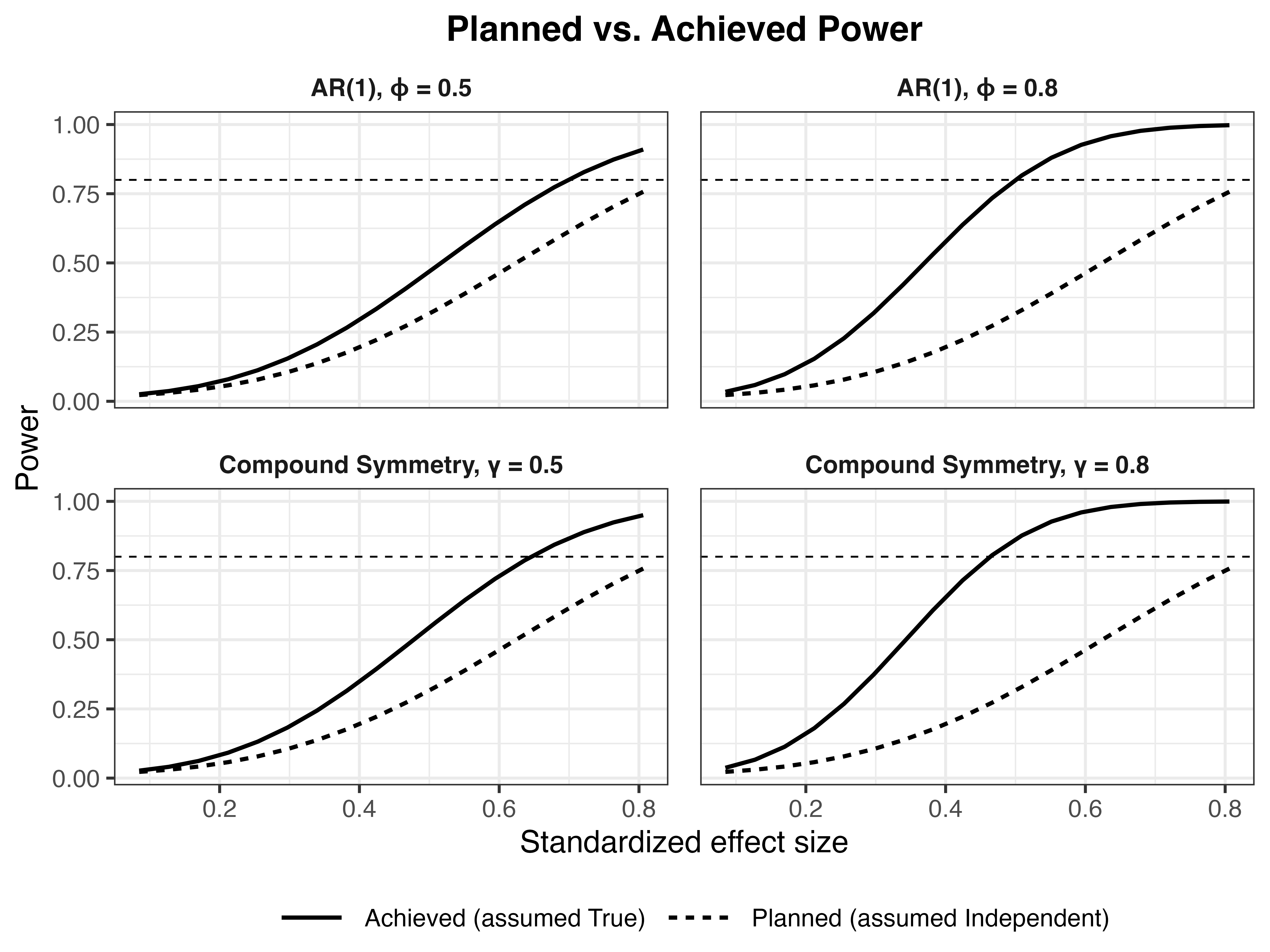}
    \caption[Planned Vs Achieved Power Under Design Stage Correlation Misspecification]{Analytic power curves across standardized effect sizes for a fixed multiple-intervention stepped wedge design. The dashed curves show planned power calculated under an independence assumption at the design stage, while the solid curves show achieved power under the true cluster-period correlation structure indicated in each panel (AR(1) with $\phi \in \{0.5,0.8\}$ and compound symmetry with $\gamma \in \{0.5,0.8\}$. The horizontal dashed line denotes the typical 80\% power threshold. Results for an 8 cluster, 6 period, symmetric factorial repeated-cross section design with $n=50$ individuals per cluster-period, $\rho_b=0.05$ and $\rho_w=0.2$.}
    \label{fig:aim1planachiev}
\end{figure}

Figure \ref{fig:aim1planachiev} shows \textit{planned} analytic power, computed under an independence working assumption at the design phase and \textit{achieved} analytic power under the true cluster-period correlation structure. Across all correlation structures considered, achieved power under the true model exceeded planned power under independence for a fixed effect size. Thus, for the M-SWDs examined here, independence-based planning is conservative such that power calculations that ignore cluster-period correlations underestimate the efficiency of the design and consequently understate the power that would be achieved in practice. However, this apparent increase in power reflects the LMM-Independent method’s failure to control Type I error and standard error deflation, rather than a genuine gain in efficiency, and should therefore be interpreted with caution.

These differences can be formalized using a generalized design effect, defined here as the ratio of the variance of the treatment effect estimator under the true cluster-period correlation structure to the variance assumed at the design stage. This definition generalizes the classical design effect used in cluster randomized trials - often represented as $1 + (N-1)\rho$ - to account for efficiency's dependence on the full cluster-period covariance structure and estimand of interest in M-SWDs. For the design examined here, design effects for main effects range from 0.30 for Compound Symmetry ($\gamma=0.8$) to 0.69 for AR(1) ($\phi=0.5$), indicating that independence-based planning overestimates the variance of the treatment effect estimator by $45-230\%$ relative to the true model. 

Conversely, assuming compound symmetry when the true correlation is weaker or independent can result in substantial variance inflation, demonstrating that misspecifying the magnitude of correlation can be as consequential as misspecifying its structure. Results for a scenario that assumes Compound Symmetry with $\gamma=0.5$ can be found in Table \ref{tab:cs_design_effect}.

\begin{table}[ht]
\centering
\caption[Design Effect Under Compound Symmetry]{Design-stage variance relative to compound symmetry planning ($\gamma=0.5$). Design effects are defined as the ratio of the variance of the main effect estimator under the true correlation structure to the variance assumed at the design stage under compound symmetry with $\gamma=0.5$.}
\label{tab:cs_design_effect}
\begin{tabular}{l c}
\hline \\
True correlation structure & Design effect ($\mathrm{Var}_{\text{true}} / \mathrm{Var}_{\text{CS}(\gamma=0.5)}$) \\ \\
\hline \\
AR(1), $\phi = 0.5$        & 1.17 \\ \\
AR(1), $\phi = 0.8$        & 0.60 \\ \\
CS, $\gamma = 0.8$         & 0.52 \\ \\
Independent                & 1.71 \\ 
\hline
\end{tabular}
\end{table}

These results show that the dominant design-stage risk in multiple-intervention stepped wedge trials is failing to account for cluster-period correlation. While independence-based planning may be conservative for certain M-SWD contrasts, assuming a correlated structure such as compound symmetry does not uniformly protect against design-stage misspecification. Accordingly, realistic assumptions about both the presence and form of cluster-period correlation should be incorporated into the planning of M-SWDs to ensure that nominal power targets translate into reliable operating characteristics.

\section{Discussion}\label{sec:discussion1}

We developed a generalized framework for modeling multiple-intervention stepped wedge designs that allows for structured within-cluster correlation and a unified representation of the intra-cluster and cluster-period dependence. While prior work has emphasized the role of intracluster correlation in stepped wedge designs, our results demonstrate that, in multiple-intervention settings, efficiency and power are governed more directly by the correlation structure of cluster-period means. By expressing treatment effect variances in terms of the standardized cluster-period mean correlation matrix $\Omega=\{\omega_{jm}\}$, we showed that assumptions about temporal dependence determine which cluster-period contrasts contribute meaningful information to estimation. As a result, commonly adopted exchangeable or independence assumptions can substantially mischaracterize efficiency in realistic designs, particularly for interaction effects and contrasts spanning distant periods. Our expansion of existing methodology enables the growing multiple-intervention stepped wedge design to be more robustly aligned with the temporal correlation structures likely to arise in pragmatic implementation settings.

 Misspecification of the cluster-period correlation structure was found to have systematic impact on standard error estimation, and consequently on the relationship between nominal and achieved power, even when the rest of the model was correctly specified. When data are generated under distance-dependent correlation structures but analyzed assuming independence or even exchangeability, power was consistently overestimated relative to the correctly specified analyses. This behavior arises because misspecifying $R$ induces an incorrect cluster-period mean correlation structure $\Omega$, leading to inappropriate weighting of cluster-period contrasts in estimation. In particular, assuming independence assigns equal weight to all contrasts, including those spanning distant periods that are weakly informative under true correlation structure. Under correct specification of the mean model, misspecification does not induce bias in point estimates, but instead distorts standard error estimation through incorrect weighting of cluster-period contrasts.  

These findings have distinct implications for design-stage planning and analysis-stage implementation. At the analysis stage, linear mixed models with compound symmetry were found to perform robustly across a range of true correlation structures, yielding stable inference with respect to variance estimation and hypothesis testing even when temporal dependencies deviated from strict exchangeability. In contrast, at the design stage, assumptions about the cluster-period correlation structure play a more critical role. Power calculations based on independence or exchangeability can be optimistic when temporal correlation decays with distance, particularly for interaction effects. Even within compound symmetry, incorrect assumptions about correlation parameter magnitude ($\gamma$) can lead to significant design effect inflation or deflation, with corresponding over- or underestimation of required sample size. Consequently, realistic design-stage planning for multiple-intervention stepped wedge trials should utilize available data and attempt to account for plausible temporal correlation structures rather than relying solely on default assumptions. 

The results of this work were derived under a set of simplifying assumptions chosen to isolate the role of cluster-period correlation in multiple-intervention stepped wedge designs. Specifically, we focus on symmetric designs with balanced cluster-period sample sizes and complete observation schedules. We also consider continuous outcomes using linear mixed-effects models with identity links. Treatment effects are assumed to be constant once implemented and the fixed-effects mean model is correctly specified. Under these conditions, correlation misspecification primarily affects variance estimation and power rather than point estimation. While these assumptions allow us to best isolate the mechanisms by which correlation structure governs efficiency, extensions to asymmetric or incomplete designs, unequal cluster sizes, non-normal outcomes, or settings with treatment heterogeneity are beyond the scope of this work. 


\bibliographystyle{unsrtnat}
\bibliography{References}

\pagebreak

\section{Supplemental Materials}

\subsection{Supplemental Materials: Variance Estimation Under Distance-Dependent Correlation Structures}
\label{sect:var-ddcorr}
When $R$ follows a distance-dependent correlation structure such as AR(1), proportional decay, or Toeplitz, the cluster-period covariance matrix, $\mathbf{V}_i = \sigma^2_\nu \mathbf{R}+ \sigma^2_c\mathbf{I}_T+(\sigma^2_\alpha + \sigma^2_\zeta)\mathbf{J}_T $ can be inverted using the Usmani Inversion \cite{usmani_inversion_1994, da_fonseca_eigenvalues_2007} and Sherman-Morrison rank one update \cite{sherman_adjustment_1950, bartlett_inverse_1951}. Here we derive the variance under the AR(1) structure as a representative example, however, the proportional decay and Toeplitz follow using similar principles. 

Under AR(1),
\begin{equation*}
    \mathbf{V}_i=\sigma^2_\nu \mathbf{R}_{\text{AR(1)}}+ \sigma^2_c\mathbf{I}_T+(\sigma^2_\alpha + \sigma^2_\zeta)\mathbf{J}_T 
\end{equation*}
where
\begin{equation*}
\mathbf{R}_{\text{AR(1)}} = \begin{pmatrix}
1 & \phi & \phi^2 & \ldots & \phi^{T-1} \\
\phi & 1 & \phi & \ldots & \phi^{T-2} \\
\phi^2 & \phi & 1 & \ldots & \phi^{T-3} \\
\vdots & \vdots & \vdots & \ddots & \vdots \\
\phi^{T-1} & \phi^{T-2} & \phi^{T-3} & \ldots & 1
\end{pmatrix}
\end{equation*}
for $\phi \in (0,1)$ such that $r_{jm}=\phi^{|j-m|}$. We invert $\mathbf{R}_{\text{AR(1)}}$
\begin{equation*}
[\mathbf{R}_{\text{AR(1)}}]^{-1} = \mathbf{Q} = \begin{pmatrix}
\frac{1}{1-\phi^2} & \frac{-\phi}{1-\phi^2} & 0 & \ldots & 0 \\
\frac{-\phi}{1-\phi^2}  & \frac{\phi^2+1}{1-\phi^2}  & \frac{-\phi}{1-\phi^2}  & \ldots & 0\\
0 & \frac{-\phi}{1-\phi^2}  & \frac{\phi^2+1}{1-\phi^2}  & \ldots & 0 \\
\vdots & \vdots & \vdots & \ddots & \vdots \\
0 & 0 & 0 & \ldots & \frac{1}{1-\phi^2} 
\end{pmatrix},
\end{equation*}
which is tridiagonal. Since $\mathbf{I}=\mathbf{QR}$, we can say 
\begin{equation*}
    \mathbf{V}=\sigma^2_\nu \mathbf{R} + \sigma^2_c\mathbf{QR} + (\sigma^2_\alpha + \sigma^2_\zeta)\mathbf{J}_T = (\sigma^2_\nu \mathbf{I}+ \sigma^2_c\mathbf{Q})\mathbf{R} + (\sigma^2_\alpha + \sigma^2_\zeta)\mathbf{J}_T. 
\end{equation*}
We define $\mathbf{W}=(\sigma^2_\nu \mathbf{I} + \sigma^2_c\mathbf{Q})\mathbf{R}$. Then we can say $\mathbf{W}^{-1}=\mathbf{Q}(\sigma^2_\nu \mathbf{I} + \sigma^2_c\mathbf{Q})^{-1}$. We can define $\mathbf{A}=\sigma^2_\nu \mathbf{I} + \sigma^2_c\mathbf{Q}$ and write $\mathbf{W}=\mathbf{AR}$ and $\mathbf{W}^{-1}=\mathbf{QA}^{-1}$. To solve for $\mathbf{A}$,

\begin{align*}
    \mathbf{A} = \sigma^2_\nu \mathbf{I} + \sigma^2_c \mathbf{Q}  & = \begin{pmatrix}
        \sigma^2_\nu & 0 & 0 & \ldots & 0 \\
        0 & \sigma^2_\nu & 0 &  \ldots & 0 \\
        0 & 0 & \sigma^2_\nu & \ldots & 0 \\
        \vdots & \vdots & \vdots & \ddots & \vdots \\
        0 & 0 & 0 & \ldots & \sigma^2_\nu
    \end{pmatrix} 
    + \begin{pmatrix}
\frac{\sigma^2_c}{1-\phi^2} & \frac{-\sigma^2_c\phi}{1-\phi^2} & 0 & \ldots & 0 \\
\frac{-\sigma^2_c\phi}{1-\phi^2}  & \frac{\sigma^2_c(1+\phi^2)}{1-\phi^2}  & \frac{-\sigma^2_c\phi}{1-\phi^2}  & \ldots & 0\\
0 & \frac{-\sigma^2_c\phi}{1-\phi^2}  & \frac{\sigma^2_c(1+\phi^2)}{1-\phi^2}  & \ldots & 0 \\
\vdots & \vdots & \vdots & \ddots & \vdots \\
0 & 0 & 0 & \ldots & \frac{\sigma^2_c}{1-\phi^2} 
\end{pmatrix}
\\ & = \begin{pmatrix}
    \sigma^2_\nu + \frac{\sigma^2_c}{1-\phi^2} & \frac{-\sigma^2_c\phi}{1-\phi^2} & 0 & \ldots & 0 \\
\frac{-\sigma^2_c\phi}{1-\phi^2}  & \sigma^2_\nu +\frac{\sigma^2_c(1+\phi^2)}{1-\phi^2}  & \frac{-\sigma^2_c\phi}{1-\phi^2}  & \ldots & 0\\
0 & \frac{-\sigma^2_c\phi}{1-\phi^2}  & \sigma^2_\nu + \frac{\sigma^2_c(1+\phi^2)}{1-\phi^2}  & \ldots & 0 \\
\vdots & \vdots & \vdots & \ddots & \vdots \\
0 & 0 & 0 & \ldots & \sigma^2_\nu + \frac{\sigma^2_c}{1-\phi^2} 
\end{pmatrix}.
\end{align*}
Thus $\mathbf{A}$ is a symmetric tridiagonal matrix and can be inverted following the Usmani Inversion \cite{usmani_inversion_1994, da_fonseca_eigenvalues_2007}. Following the format from Usmani, we can break down the $\mathbf{W} = [w_{jm}]$ matrix, for $j,m = 1, \ldots, T$ such that
\begin{align*}
    & w_{11} = w_{TT} = \sigma^2_\nu + \frac{\sigma^2_c}{1-\phi^2}, & \hspace{4pt}  & w_{jj} = \sigma^2_\nu + \frac{\sigma^2_c (\phi^2+1)}{1-\phi^2} \\
    & w_{j,j+1} = w_{j,j-1}=\frac{-\sigma^2_c \phi}{1-\phi^2}, & \hspace{4pt} & 0 \text{ otherwise} 
\end{align*}
and we can write the principal minors
\begin{align*}
    & a_1=a_T = w_{11}=w_{TT}, & & a_j=w_{jj} \text{ for } j \in (2 \leq j \leq T-1), & &  \\ &  b_j=w_{j,j+1} & &c_j=w_{j,j-1}, \\
    & g_j=a_jg_{j-1}-b_{j-1}c_{j-1}g_{j-2} \text{ for } j=\{2, 3, \ldots, T\}, & & g_{0}=1, g_1=a_1 ,
\end{align*}
and the recurrence formula 
\begin{align*}
    h_j=a_jh_{j+1}-b_jc_jh_{j+2} \hspace{4pt} \text{  for  } j=\{T-1, T-2, \ldots, 2, 1\}, & & h_{T+1}=1, & & h_{T}=a_T. 
\end{align*}
Following Usmani, we can invert this framework such that $\mathbf{A}_T^{-1}=[s_{jm}]$ with 
\begin{equation*}
    s_{jm} = \begin{cases}
        (-1)^{j+m}c_jc_{j+1}\ldots c_{m-1}g_{j-1}h_{m+1}/g_T & j <m \\ 
        g_{j-1}h_{j+1}/g_T & j=m \\
        (-1)^{j+m}b_{m+1}b_{m+2}\ldots b_{j-1}g_{m-1}h_{j+1}/g_T & j > m
    \end{cases} .
\end{equation*}
Since $b_j=c_j=b=-\frac{\sigma^2_c\phi}{1-\phi^2}$ for all $j$ due to consistent first off-diagonals we can simplify $c_jc_{j+1}\ldots c_{m-1}= b^{j-m}$ and $b_{m+1}b_{m+2}\ldots b_j=b^{m-j}$. Since $(j+m)$ and $(m-j)$ have the same parity, we can write: $(-1)^{j+m}b^{m-j}=(-b)^{m-j}$. Thus
\begin{equation*}
    s_{jm} = \begin{cases}
        (-b)^{m-j}g_{j-1}h_{m+1}/g_T & j < m\\
              g_{j-1}h_{j+1}/g_T & j=m \\
        (-b)^{j-m}g_{m-1}h_{j+1}/g_T & j > m
    \end{cases} .
\end{equation*}
We can further unify this:
\begin{equation*}
    s_{jm} = (-b)^{|j-m|}g_{\text{min}(j,m)-1}h_{\text{max}(j,m)+1}/g_T  \hspace{10pt} \text{ for all } j,m \in \{1, 2, \ldots, T\}.
\end{equation*}


Now that we've solved for $\mathbf{A}^{-1}$ and $\mathbf{Q}$ we recall that $\mathbf{V}=\mathbf{W}+(\sigma^2_\alpha + \sigma^2_\zeta)\mathbf{J}_T$ with $\mathbf{W}^{-1}=\mathbf{QA}^{-1}$. To finish solving for the inverse of $\mathbf{V}$, we can write $\mathbf{V}=\mathbf{W}+k\mathbf{11}'$ where $k=(\sigma^2_\alpha + \sigma^2_\zeta)$ and $\mathbf{1}\in \mathbb{R}$ is a vector of ones. Under this parameterization, we can use the Sherman-Morrison rank one update \cite{sherman_adjustment_1950, bartlett_inverse_1951} such that:
\begin{equation*}
    \mathbf{V}^{-1}=(\mathbf{W}+uv')^{-1} \text{ with } u=v=\sqrt{k}\mathbf{1}.
\end{equation*}
Then, following the rank one update,
\begin{align*}
    \mathbf{V}^{-1}&=(\mathbf{W}+uv')^{-1} =\mathbf{W}^{-1}-\mathbf{W}^{-1}u(1+v'\mathbf{W}^{-1}u)^{-1}v'\mathbf{W}^{-1}\\
    &=\mathbf{W}^{-1}-\mathbf{W}^{-1}(\sqrt{k}\mathbf{1})(1+k\mathbf{1}\mathbf{W}^{-1}\mathbf{1})^{-1}(\sqrt{k}\mathbf{1})'\mathbf{W}^{-1}
\end{align*}
Let $\mathbf{d}=\mathbf{W}^{-1}\mathbf{1}$ and $\mathbf{p}=\mathbf{1}'\mathbf{d}$. Thus $(\sqrt{k}\mathbf{1})'\mathbf{W}^{-1}(\sqrt{k}\mathbf{1})=k\mathbf{p}$.

Finally we can say:
\begin{equation}
    \mathbf{V}^{-1}=\mathbf{W}^{-1}-\frac{k}{1+k\mathbf{p}}\mathbf{dd}'.
\end{equation}

\subsection{Supplemental Materials: Analysis Methods}

We presented results addressing three related questions: (i) whether the analytical power formulas for compound symmetry and AR(1) derived in the main and supporting text can be validated empirically; (ii) how correlation structure affects power under correct specification; and (iii) how misspecifying the working covariance affects inference and design choices.

To evaluate these questions, we considered four complementary analysis methods. Method 1 was a closed form theoretical generalized least squares (GLS)-based power calculation utilizing the findings in the main and supporting text. Method 2 was simulation-based GLS "oracle" estimator using the true covariance structure. Method 3 was a linear mixed model assuming compound symmetry (LMM-CS). Method 4 was a linear mixed model assuming independence (LMM-Independent). Together these approaches span the spectrum from idealized efficiency bounds to realistic applied analyses and allow isolation of the effects of covariance specification from sampling variability and model estimation error. 

\noindent \textbf{Analytic (Closed-Form GLS):} Analytic power calculations were obtained using the closed-form variance expressions derived in the main and supporting text. These calculations treated the cluster-period means as the unit of analysis and assumed that the true covariance matrix of the cluster-period means, $\mathbf{V}$, is known and correctly specified. For a given design matrix $\mathbf{Z}$ and covariance matrix $\mathbf{V}$, the variance of the GLS estimator was
\begin{equation*}
    \text{Var}(\boldsymbol{\hat{\theta}})=(\mathbf{Z'V}^{-1}\mathbf{Z})^{-1}
\end{equation*}
from which standard errors were obtained as the square root of the diagonal elements ($\sqrt{\text{Var}(\hat{\theta}_q)} = \sqrt{[(\mathbf{Z}'\mathbf{V}^{-1}\mathbf{Z})^{-1}]_{qq}}$) for the $q$th element of the parameter vector $\boldsymbol{\theta}$. 

Power was computed using a normal approximation to the Wald test. For a true effect size, $\theta_q$, two-sided tests with significance level $\alpha$ (adjusted for multiplicity using Bonferroni corrections where appropriate) were evaluated as
\begin{equation*}
    \text{Power} = \Phi(|\frac{\theta_{q,a}}{\sqrt{\text{Var}(\hat{\theta_q})}}| - z_{\alpha/2})
\end{equation*}
where $\Phi(\cdot)$ denotes the standard normal cumulative distribution function. These calculations required no simulations and provide a computationally efficient benchmark that supported the methodological findings of this work, while representing the best-cases scenario in which the true covariance structure is known. 

\noindent \textbf{GLS-Oracle (Simulation-Based):} GLS-Oracle estimation was used to empirically validate the analytical variance expressions and to isolate the effects of finite-sample variability while retaining perfect knowledge of the true covariance structure. Data were generated under the full individual-level model and subsequently collapsed to cluster-period means. For each simulated dataset, generalized least squares estimation was performed using the true cluster-period covariance matrix $\mathbf{V}$ used in data generation. Specifically, for intervention $q$
\begin{equation*}
    \hat{\theta}_q=(\mathbf{Z'V}^{-1}\mathbf{Z})^{-1}\mathbf{Z'V}^{-1}\bar{Y}^{(q)}
\end{equation*}
was used for treatment effect estimates and model-based standard errors. Hypothesis testing was conducted using Wald tests with Bonferroni-adjusted significance levels. 

Unlike the analytical approach, the GLS-Oracle method incorporated Monte Carlo variability arising from outcome generation while eliminating uncertainty due to covariance estimation or misspecification. As such, it represents an empirical, albeit unobtainable in practice, gold standard against which both analytical calculations and mixed-model approaches can be compared. 

\noindent \textbf{Linear Mixed Model with Compound Symmetry (LMM-CS):} Linear mixed models assuming compound symmetry (LMM-CS) were fit to the individual-level data using restricted maximum likelihood (REML). LMM-CS models included a fixed effect for time, a random intercept for clusters, and a random cluster-by-period effect corresponding to an exchangeable correlation structure at the cluster-period level. The fitted model was:
\begin{equation*}
    Y_{ijk} = \mu + \beta_j+ \alpha_i + \nu_{ij} + \psi_{ik}+\theta_1X_{1ij} + \theta_2X_{2ij}+\theta_3X_{1ij}X_{2ij} + e_{ijk}
\end{equation*}
with $\alpha_i \sim N(0,\sigma^2_\alpha), \nu_{ij}\sim N(0,\sigma^2_\nu), \psi_{ik}\sim N(0,\sigma^2_\psi), \text{ and } e_{ijk}\sim N(0,\sigma^2_e)$ and $\sigma^2_\psi=0$ under repeated cross-sectional designs. 

Under this specification, the implied cluster-period covariance assumed constant correlation between all periods within a cluster, regardless of temporal separation, similar to compound symmetry. Model-based standard errors were extracted from the fitted variance-covariance matrix of fixed effects. Wald tests were used for inference with Bonferroni correction applied to account for multiple comparisons. 

This approach reflects common practice in applied stepped wedge analysis and represents a realistic scenario in which cluster-period correlation is modeled but potentially misspecified. 

\noindent \textbf{Linear Mixed Model with Independence (LMM-Independent):} Linear mixed models assuming independence (LMM-Independent) were fit to the individual-level data. These models included a fixed effect for time but only a cluster-level random intercept while omitting the cluster-period random effect:
\begin{equation*}
    Y_{ijk} = \mu + \beta_j+ \alpha_i + \psi_{ik}+\theta_1X_{1ij} + \theta_2X_{2ij}+\theta_3X_{1ij}X_{2ij} + e_{ijk}.
\end{equation*}
This specification assumed independence of observations across time within clusters beyond the persistent cluster effect and ignored any additional correlation induced by repeated measurement of cluster-period means. 

LMM-Independent serves as a deliberately misspecified baseline, representing analyses that fail to account for within-cluster temporal correlation. Comparing its performance to LMM-CS and GLS-based methods provides insight into the consequences of ignoring cluster-period correlation in multiple-intervention stepped wedge designs.

\subsection{Supplemental Materials: Asymptotic Power Validation}
Under asymptotic power validation, we find small average differences (average difference [$\text{mean}(\text{Power}_\text{Simulated} -\text{Power}_{\text{Theoretical}})$] $<0.001$; average absolute difference [$\text{mean}(|\text{Power}_\text{Simulated} -\text{Power}_{\text{Theoretical}}|)$] $0.003$), with a maximum absolute difference of $0.012$. Percent differences [$\frac{\text{Power}_\text{Simulated}-\text{Power}_{\text{Theoretical}}}{\text{Power}_{\text{Theoretical}}}*100$] center near zero (mean = $0.167$) with a range from $-12.0$ to $14.5$. As expected, larger percent deviations occurred primarily where power was near zero, while absolute differences remained negligible across the full power range.

\begin{figure}
    \centering
    \includegraphics[width=0.8\linewidth]{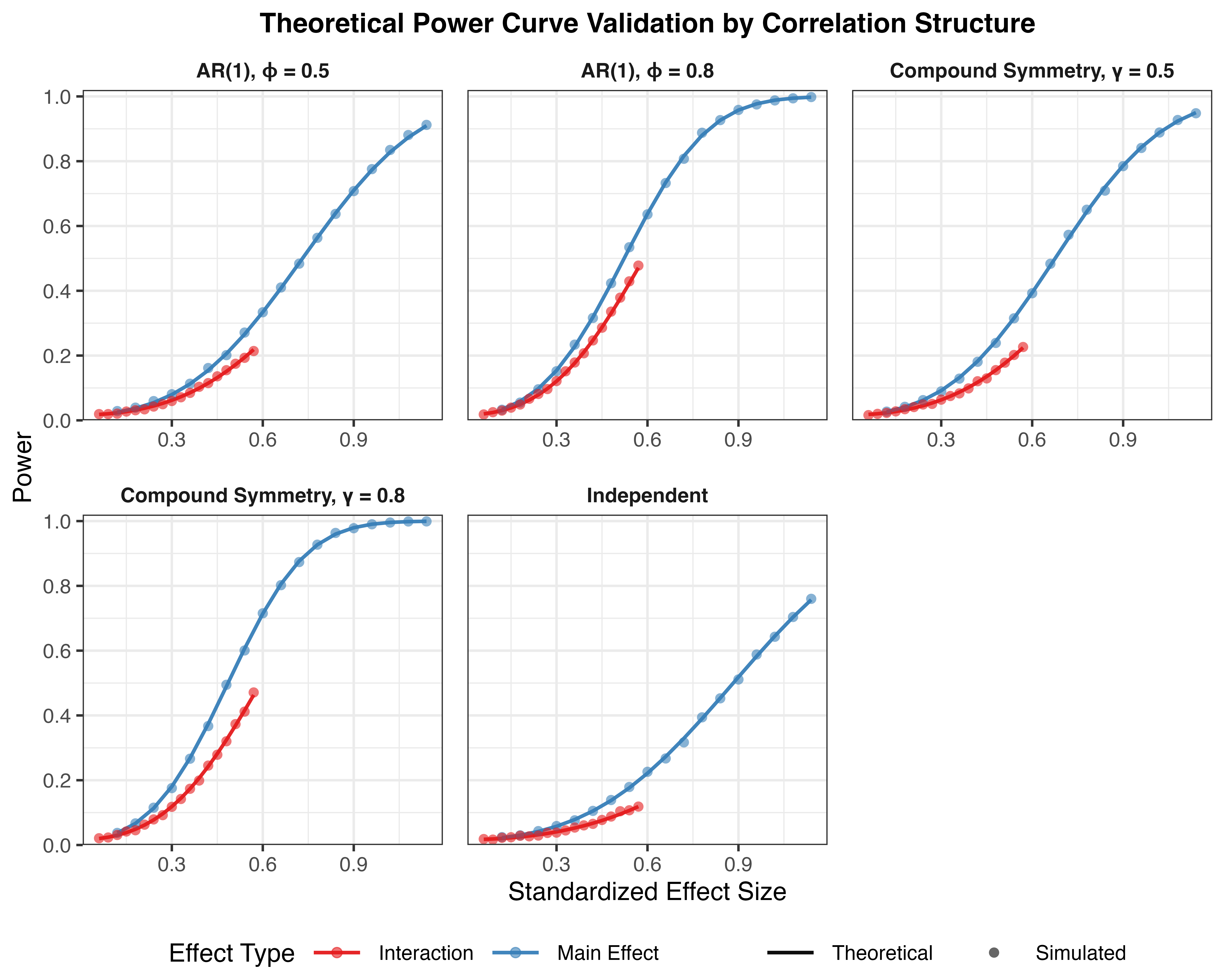}
    \caption[Theoretical Power Curves for Validation of Correlation Structure Methodology]{Comparison of theoretical power curves and simulated power curves. Results for an 8 cluster, 6 period, symmetric factorial repeated-cross section design with $n=50$ individuals per cluster period, $\rho_b=0.05$ and $\rho_w=0.2$. Solid lines represent theoretical power and points represent simulated power. Blue lines and points are main effects and red lines and points are interaction effects. Due to the symmetry of main effects (X1 and X2), results are collapsed into X1. Details for X2 can be found in the appendix.}
    \label{fig:aim1validation}
\end{figure}

\begin{figure}
    \centering
    \includegraphics[width=0.8\linewidth]{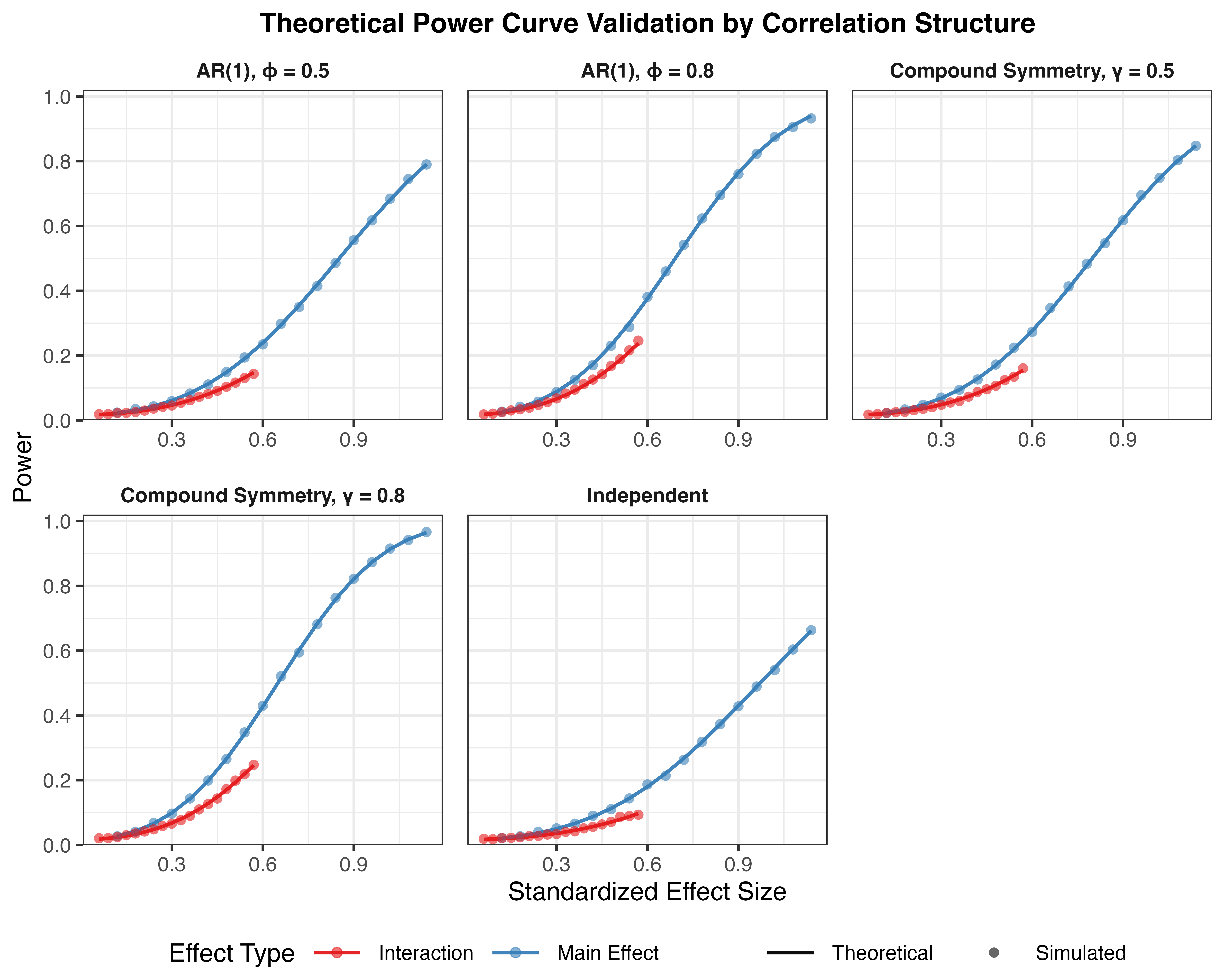}
    \caption[Theoretical Power Curves for Validation of Correlation Structure Methodology - Small Sample Size]{Comparison of theoretical power curves and simulated power curves under small sample size ($n=15$). Results for an 8 cluster, 6 period, symmetric factorial repeated-cross section design with $\rho_b=0.05$ and $\rho_w=0.2$. Solid lines represent theoretical power and points represent simulated power. Blue lines and points are main effects and red lines and points are interaction effects. Due to the symmetry of main effects (X1 and X2), results are collapsed into X1.}
    \label{fig:aim1validation_2}
\end{figure}

\begin{figure}
    \centering
    \includegraphics[width=0.8\linewidth]{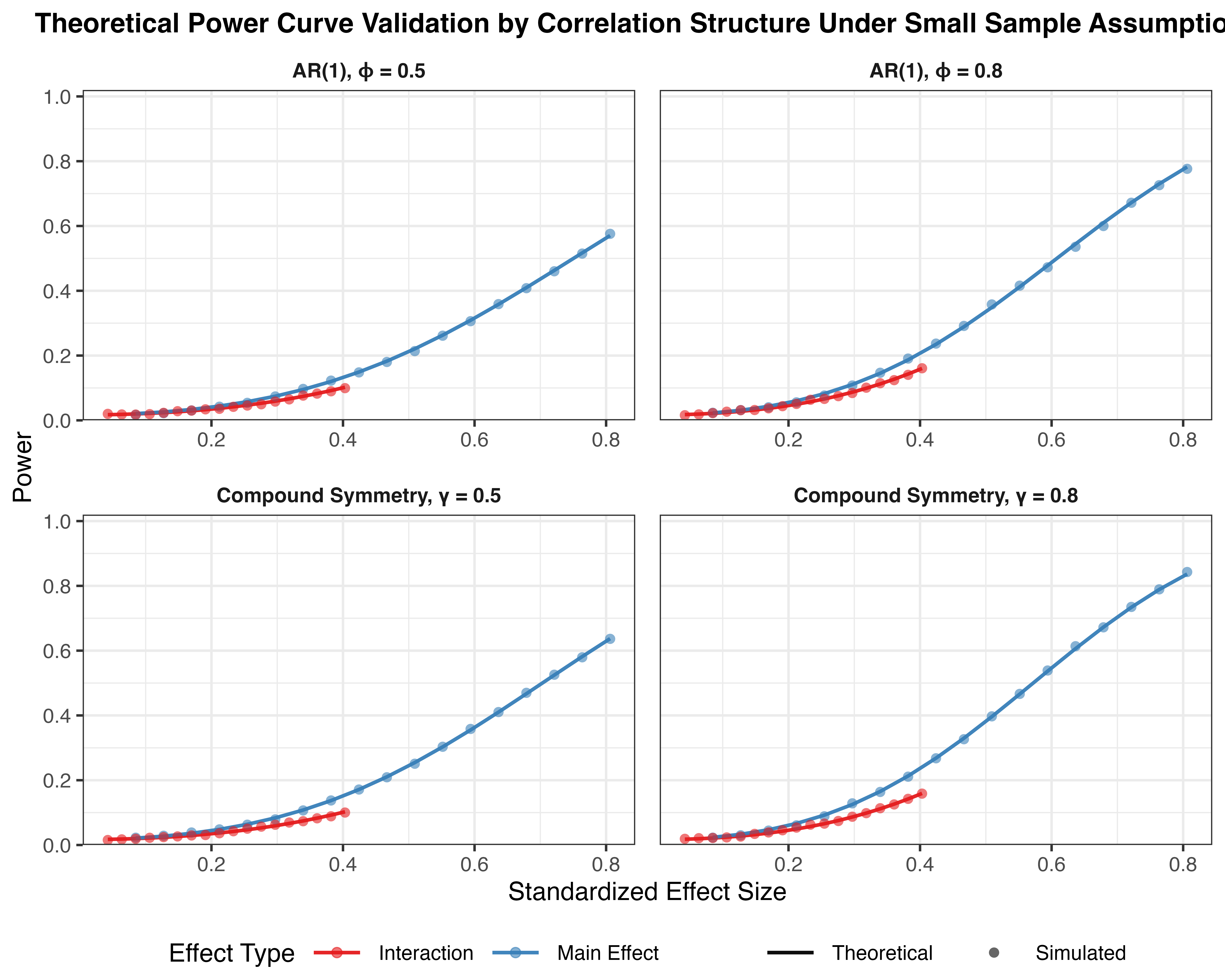}
    \caption[Theoretical Power Curves for Validation of Correlation Structure Methodology - Small Sample Size]{Comparison of theoretical power curves and simulated power curves under small sample size ($n=15$) and fewer clusters ($I=6$). Results for a 6 period symmetric factorial repeated-cross section design with $\rho_b=0.05$ and $\rho_w=0.2$. Solid lines represent theoretical power and points represent simulated power. Blue lines and points are main effects and red lines and points are interaction effects. Due to the symmetry of main effects (X1 and X2), results are collapsed into X1.}
    \label{fig:aim1validfew}
\end{figure}

\begin{small}
\begin{table}[]
    \caption[Table of Correlation Structure Validation Results Under Multiple Scenarios]{Table of Average Difference (Avg. Dif.), Average Absolute Difference (Avg. Abs. Dif.), Maximum Absolute Difference (Max. Abs. Dif.), Mean Percentage Difference (Mean $\%$ Dif.), and Range for 6 scenarios. Scenarios indicate whether the design was factorial (F) or concurrent (C), the value of the ICCs ($\rho_w$ and $\rho_a$), cluster-period sample size ($n$), and number of clusters ($I$).}
    \label{tab:aim1validmore}
    \centering
    \begin{tabular}{c c c c c c c}
   \hline \\
        Scenario(F/C, $\rho_w,\rho_b,n, I$) & Avg.Dif. & Avg.Abs.Dif & Max.Abs.Dif. & Mean $\%$Dif. & $\%$Range  \\
         \hline \\
         \textbf{Base (F, 0.2, 0.05, 50, 8)} & & & & & & \\
        Overall & $<$0.001 & 0.003 & 0.012 & 0.167 & (-12.0,14.5) \\
        X1 & $<$0.001 & 0.003 & 0.012 & 0.257 & (-6.8, 12.8) \\
        X2 & $<$0.001 & 0.003 & 0.010 & 0.074 & (-5.6,11.2) \\
        X1:X2 & $<$0.001 & 0.002 & 0.009 & 0.170 & (-12.0, 14.5) \\
        \hline \\
        \textbf{S1 (F, 0.2, 0.05, 15, 8)} & & & & & \\
       Overall & $<$0.001 & 0.003 & 0.012 & 0.486 & (-7.8,16.1) \\
       X1 & $<$0.001 &  0.003 & 0.012 & 1.091 & (-6.4, 10.3) \\
       X2 & $<$0.001 & 0.003 & 0.011 & 0.133 & (-6.2, 16.1) \\
       X1:X2 & $<$0.001 & 0.002 & 0.008 & 0.235 & (-7.4, 13.3) \\
       \hline \\
       \textbf{S2 (F, 0.2, 0.1, 50, 8)} & & & & & \\
       Overall & $<$0.001 & 0.002 & 0.013 & 0.151 & (-12.0,15.3) \\
       X1 & $<$0.001 & 0.002 & 0.013 & 0.238 & (-4.5, 15.3) \\
       X2 & $<$0.001 & 0.003 & 0.011 & 0.018 & (-4.3, 9.0) \\
       X1:X2 & $<$0.001 & 0.002 & 0.009 & 0.198 & (-12.0, 10.0) \\
       \hline \\
       \textbf{S3 (C, 0.2, 0.05, 50, 8 )} & & & & & \\
       Overall & $<$0.001 & 0.003 & 0.014 & -0.090 & (-17.8, 14.6) \\
       X1 & $<$0.001 &  0.003 & 0.012 & 0.131 & (-13.1, 10.4) \\
       X2 & $<$0.001 & 0.003 & 0.014 & -0.212 & (-9.2, 11.1) \\
       X1:X2 & $<$0.001 & 0.002 & 0.007 & -0.190 & (-17.8, 14.6) \\
           \hline \\
       \textbf{S4 (F, 0.4, 0.3, 15, 8 )} & & & & & \\
       Overall & $<$0.001 & 0.003 & 0.013 & -0.234 & (-17.8, 14.6) \\
       X1 & $<$0.001 &  0.003 & 0.012 & -0.181 & (-7.9,9.4) \\
       X2 & $<$0.001 & 0.003 & 0.013 & -0.011 & (-6.1,14.1) \\
       X1:X2 & $<$0.001 & 0.003 & 0.011 & -0.523 & (-11.4,15.2)\\
           \hline \\
       \textbf{S5 (F, 0.2, 0.05, 15, 6 )} & & & & & \\
       Overall & $<$0.001 & 0.003 & 0.013 & -0.271 & (-18.4, 14.8) \\
       X1 & $<$0.001 &  0.003 & 0.010 & -1.330 & (-18.4,6.8) \\
       X2 & $<$0.001 & 0.003 & 0.013 & -0.098 & (-11.7,14.8) \\
       X1:X2 & $<$0.001 & 0.002 & 0.006 & 0.615 & (-12.0,7.0) \\
       \hline
    \end{tabular}

\end{table}    
\end{small}

\subsection{Supplemental Materials: Interaction Between Sample Size ($n$) and Cluster-Period Correlation Structure, $R$}

\begin{figure}
    \centering
    \includegraphics[width=0.8\linewidth]{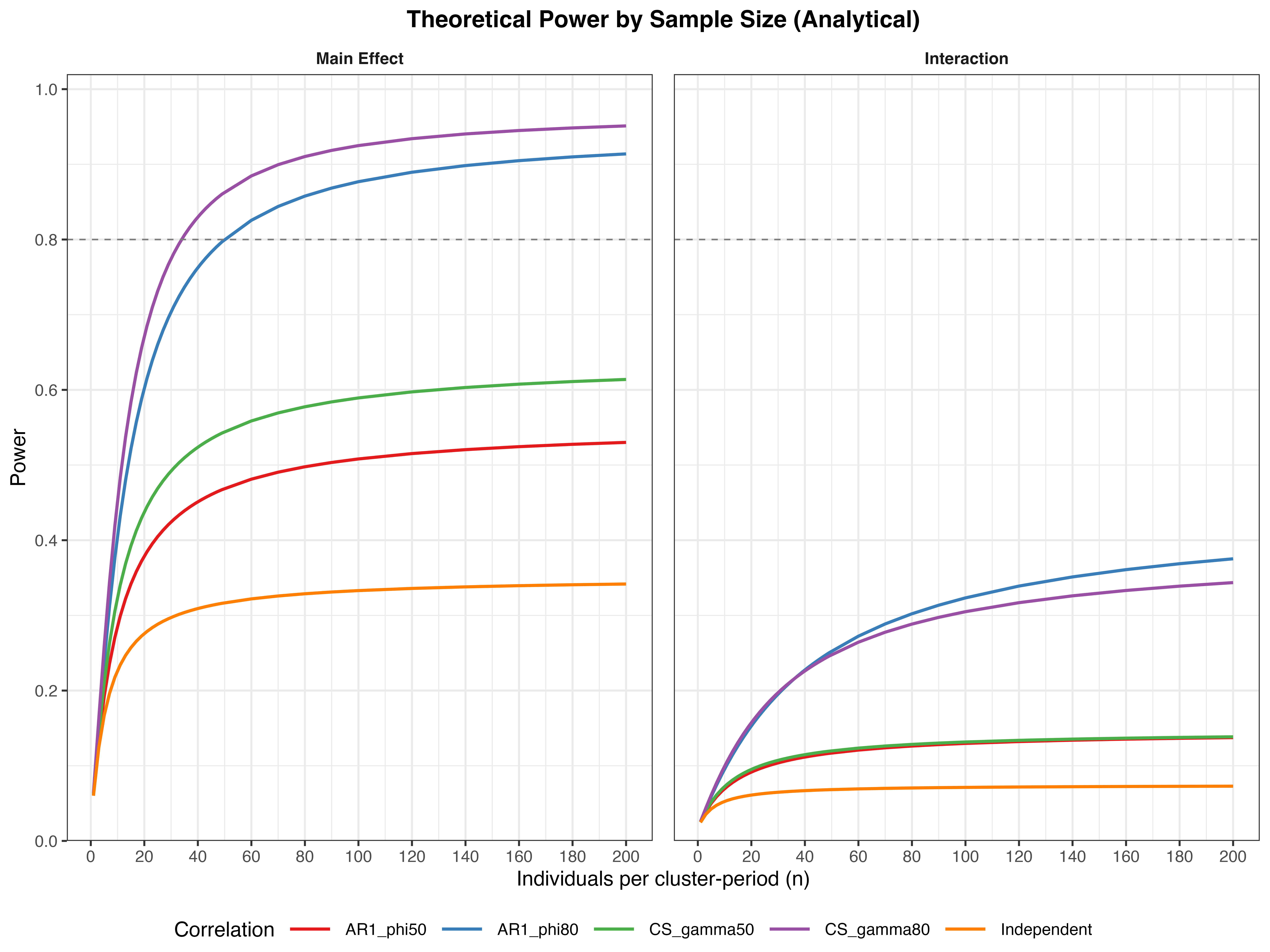}
    \caption[Sensitivity of Power to Sample Size Under Correlated Structures]{Power curves shown for main effects (left) and interaction effect (right) across number of individuals per cluster-period ($n$) are computed using closed-form variance expressions under correct specification of the correlation structure $R$. Results for an 8 cluster, 6 period, symmetric factorial repeated-cross section design with $\rho_b=0.05$, $\rho_w=0.2$, Cohen's $d$ standardized main effect $0.6$ and interaction effect $0.3$. Results illustrate how power gains from increasing $n$ depend on the assumed temporal correlation structure.}
    \label{fig:Aim1Q1SS}
\end{figure}

\begin{figure}
    \centering
    \includegraphics[width=0.8\linewidth]{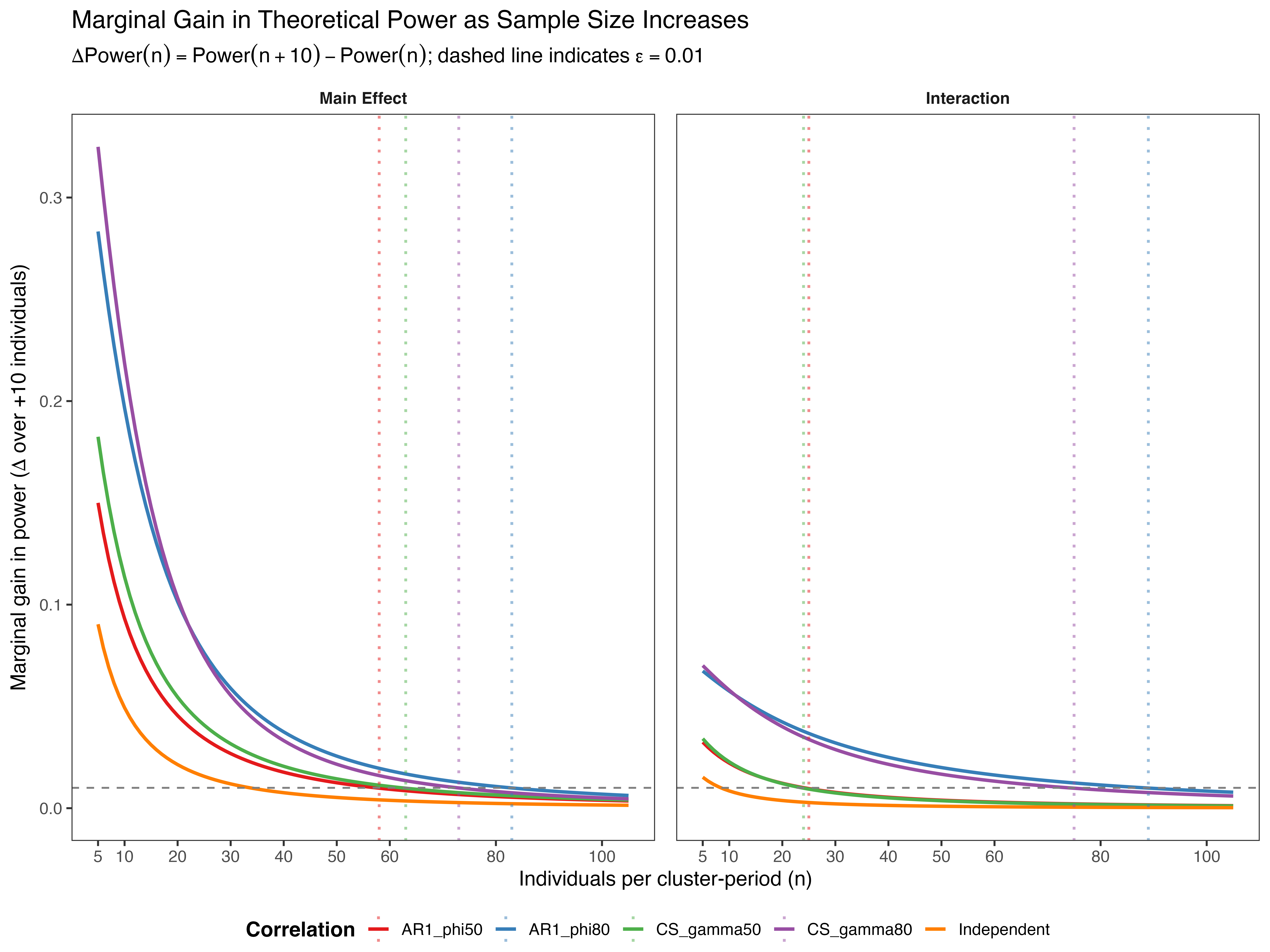}
    \caption[Marginal Gain in Theoretical Power as Sample Size Increases]{Theoretical power curves show the marginal increase in power from adding 10 individuals per cluster-period, $ \Delta\text{Power}(n) = \text{Power}(n+10)-\text{Power}(n)$ for main effects (left) and interaction effect (right). The dashed horizontal line denotes a determined negligible gain threshold ($\epsilon=0.01$). Vertical dotted lines indicate the saturation point for diminishing returns under correlated structures. Independence is shown for baseline comparison. Results for an 8 cluster, 6 period, symmetric factorial repeated-cross section design with $\rho_b=0.05$, $\rho_w=0.2$, Cohen's $d$ standardized main effect $0.6$ and interaction effect $0.3$.}
    \label{fig:Aim1Q1SSdimret}
\end{figure}

Increasing the number of individuals per cluster-period reduces the individual-level residual variance but does not affect the cluster-level or cluster-period components that induce correlation across observations. Consequently, efficiency gains from increasing sample size in M-SWDs depend on how the temporal correlation structure $R$ governs the accumulation of information across time. Once individual-level variability becomes small relative to correlated components, additional observations primarily reuse information rather than contributing independent signal. 

Figure \ref{fig:Aim1Q1SS} shows theoretical power as a function of the number of individuals per cluster-period ($n$), stratified by the cluster-period correlation structure $R$.  For main effects, power increases rapidly with $n$ under correlated structures but approaches a plateau as individual-level variance becomes negligible relative to correlated components. This saturation occurs earliest under compound symmetry and more gradually under AR(1) while independence exhibits slower, monotonic gains. Power to detect interaction effects remain uniformly low across all structures, even at relatively large sample sizes.

We define the \textit{saturation point} as the smallest $n$ such that 
\begin{equation*}
    \Delta\text{Power}(n) = \text{Power}(n+\delta)-\text{Power}(n) < \epsilon 
\end{equation*}
with $\delta=10$ and $\epsilon=0.01$. Figure \ref{fig:Aim1Q1SSdimret} illustrates the marginal gains in power as sample size increases. Across all correlation structures, marginal gains decline rapidly as $n$ increases, indicating  diminishing returns once individual-level variance is small. Saturation occurs earlier under compound symmetry than AR(1), reflecting greater information reuse across time, and interaction effects exhibit earlier and more pronounced saturation than main effects across all structures.   

These results demonstrate that the value of increasing within-period sample size in M-SWDs depends not only on the magnitude of correlation but also on its structure. While increasing sample size can improve efficiency for main effects, it cannot fully overcome structural limitations imposed by weak or rapidly decaying correlation, particularly for interaction effects.

\end{document}